\documentclass[12pt,journal]{IEEEtran}
\usepackage{amsmath}
\usepackage{graphicx}
\usepackage{mathtools}
\usepackage{amsfonts}
\usepackage{pifont}
\usepackage{amssymb}
\usepackage{epstopdf}
\usepackage{color}
\usepackage[utf8]{inputenc}
\usepackage{setspace}
\usepackage{ragged2e}
\usepackage{epsfig}
\usepackage{algorithmic}
\usepackage{textcomp}

\newtheorem{corollary}{Corollary}

\newtheorem{theorem}{Theorem}

\makeatletter
\newcommand{\vast}{\bBigg@{1.2}}
\newcommand{\Vast}{\bBigg@{2.5}}
\newcommand{\vastl}{\bBigg@{3.9}}
\newcommand{\Vastl}{\bBigg@{2.1}}
\makeatother

\onecolumn
\begin{document} 
\title{Performance Analysis of Physical Layer Security over Fluctuating Beckmann Fading Channels}
\author{\vspace{0.5 cm}\IEEEauthorblockN{Hussien Al-Hmood, \textit{Member, IEEE,} and H. S. Al-Raweshidy, \textit{Senior Member, IEEE}}}
\maketitle

\begin{abstract}
In this paper, we analyse the performance of physical layer security over Fluctuating Beckmann (FB) fading channel which is an extended model of both the $\kappa-\mu$ shadowed and the classical Beckmann distributions. Specifically, the average secrecy capacity (ASC), secure outage probability (SOP), the lower bound of SOP (SOP$^L$), and the probability of strictly positive secrecy capacity (SPSC) are derived in exact closed-form expressions using two different values of the fading parameters, namely, $m$ and $\mu$ which represent the multipath and shadowing severity impacts, respectively. Firstly, when the fading parameters are arbitrary values, the performance metrics are derived in exact closed-form in terms of the extended generalised bivariate Fox's $H$-function (EGBFHF) that has been widely implemented in the open literature. In the second case, to obtain simple mathematically tractable expressions in terms of analytic functions as well as to gain more insight on the behaviour of the physical layer security over Fluctuating Beckmann fading channel models, $m$ and $\mu$ are assumed to be integer and even numbers, respectively. The numerical results of this analysis are verified via Monte Carlo simulations.
\end{abstract}


\section{Introduction}
Shannon's information-theoretic notion of perfect secrecy has been developed by Wyner via suggesting the wiretap channel. In the notion of this channel, an eavesdropper is presented when a legitimate user, namely, Alice, communicates with the intended receiver which is called Bob [1]. Consequently, the performance of the physical layer security over different fading channel models has been widely analysed in the open literature. For instance, the probability of strictly positive secrecy capacity (SPSC), the secure outage probability (SOP), and the average secrecy capacity (ASC) when the wireless channels subject to the additive white Gaussian noise (AWGN) and Rayleigh fading channel are given in [2] and [3], respectively. In [4], the SPSC when both the main and eavesdropper channels undergo Rician fading channel is derived. The SOP and the SPSC of the physical layer using Rician and Nakagami-$m$ fading conditions for the Bob and the eavesdropper wireless channels are given in [5]. The Weibull fading channel model is used in [6] and [7] to study the SPSC and ASC, respectively.
\par Recently, many works have been implemented using various generalized fading distributions that unify most of the well-known channel models. In addition, they provide results closer to the practical data than the conventional distributions, namely, Rayleigh, Nakagami-$m$, and Nakagami-$n$. In [8], the ASC over $\kappa-\mu$ fading channel that is used to model the line-of-sight (LoS) communication environment is derived. The performance of the physical layer security in non-linear communication scenario is analysed in [9] and [10] via utilising the $\alpha-\mu$ fading condition. Moreover, the ASC, the SOP, the SOP$^L$, and the SPSC of the physical layer over $\alpha-\mu$ fading using the Fox's $H$-function channel model which is a unified framework for a variety of distributions are presented in [11]. The ASC using the $\kappa-\mu$/$\alpha-\mu$ and $\alpha-\mu$/$\kappa-\mu$ fading scenarios for the main/eavesdropper channels is given in [12]. The more generalised fading channels $\alpha-\kappa-\mu$ and $\alpha-\eta-\mu$ are used in [13] to derive the SOP$^L$ and its asymptotic value. These fading distributions are provided in a single model which is $\alpha-\eta-\kappa-\mu$ that is also used to represent both the main channel and the eavesdropper's channel of the classic Wyner’s wiretap model in [14]. 
\par The wireless may also affected by the multipath and shadowing simultaneously. Accordingly, the performance metrics of the physical layer security over composite fading channels have been also derived  by several efforts in the open literature. For example, the analysis in [15]-[17] are investigated over generalised-$K$ ($K_G$) fading channel which is a composite of Nakagami-$m$/gamma distributions using different methods. In [11] and [18], the Fisher-Snedecor $\mathcal{F}$ distribution that is proposed as an alternative approach for the $K_G$ fading condition via employing the inverse Nakagami-$m$ distribution instead of gamma model is used to derive the expression of the ASC, the SOP, the SOP$^L$, and the SPSC of the physical layer. The ASC and the SOP over $\kappa-\mu$ shadowed fading are given in [19] for integer fading parameters as well as in terms of the derivative of the incomplete moment generating function (IMGF) framework that is included a bivariate confluent hypergeometric function $\Phi_2(.)$. This channel model is also utilised in [20] to analyse the SOP$^L$ and SPSC using the exact probability density function (PDF) and the Gamma distribution as an approximate approach. However, the results in both efforts are either included double infinite series or approximated. Therefore, the authors in [21] have extensively analysed the performance of the physical layer security over $\kappa-\mu$ shadowed fading channel using exact closed-form analytic expressions for both scenarios of the values of the fading parameters.    
\par More recent, the so-called Fluctuating Beckmann (FB) fading channel has been proposed as an extended model of the $\kappa-\mu$ shadowed and the classical Beckmann distributions [22]. Thus, it includes the one-sided Gaussian, Rayleigh, Nakagami-$m$, Rician, $\kappa-\mu$, $\eta-\mu$, $\eta-\kappa$, Beckmann, Rician shadowed and the $\kappa-\mu$ shadowed distributions as special cases. Hence, the FB fading channel is more generalised than the $\kappa-\mu$ shadowed fading. In addition to [22], the aforementioned fading channel model has been utilised by only one previous work via studying the effective rate of communication system [23].        
\par Motivated by there is no work has been achieved to analyse the secrecy performance of the physical layer over FB fading channels, this paper is dedicated to investigate this analysis. Our main contributions are summarised as follows:
\begin{itemize}
\item Analysing the performance of the physical layer security when both the main and wiretap channels are subjected to FB fading channel models. In particular, novel exact closed-form mathematically tractable expressions of the ASC, SOP, SOP$^L$, and SPSC are derived.
\item When the fading parameters, namely, $\mu$ and $m$ which represent the real extension of multipath clusters and shadowing severity index, respectively are arbitrary numbers, the secrecy performance metrics are obtained in terms of the extended generalised bivariate Fox's $H$-function (EGBFHF). Although, this function is not available in the popular mathematical software packages such as MATLAB and MATHEMATICA, it has been implemented by several works such as [10] and [24].
\item To earn more insights into the behaviour of the physical layer security as well as the impact of the parameters of the FB fading model via using simple exact closed-from analytic expressions of the aforementioned performance metrics, $\mu$ and $m$ are assumed to even and integer values. Consequently, the derived results are obtained in simple mathematical functions that are presented in all software packages.  
\item From the provided literature in this work, the ASC, the SOP, the SOP$^L$, and the SPSC for some special cases of the FB fading model such as Beckmann have not been yet introduced due to the complexity of their PDF and cumulative distribution function (CDF). However, these expressions can be deduced from our derived expressions because the FB fading model is a versatile representation of many distributions such as $\kappa-\mu$ shadowed and Beckmann. 
\end{itemize}
\par \textit{Organization:} Section II is divided into two subsections. In the first subsection, the system model that is used in this work is described whereas the general and limited formats of the probability density function (PDF) and the cumulative distribution function (CDF) of FB fading channel are given in the second subsection. The ASC, the SOP, the SOP$^L$, and the SPSC for two cases of the values of $\mu$ and shadowing parameters are derived in Sections III, IV, V, and VI, respectively. Section VII explains the performance of the physical layer security over some special cases of FB fading channel. In Section VIII, the Monte Carlo simulations and numerical results are presented. Finally, some conclusions are highlighted in Section IX. 
\section{System and Channel Models}
\subsection{System Model}
Wyner's wiretap channel model has been proposed three different nodes with two wireless communication links [1]. The first link is between the transmitter and the legitimate receiver which are called Alice and Bob, respectively via main channel. Thus, Bob's channel state information (CSI) can be known by Alice. On the other side, the second wireless communication link describes the wiretap channel between the Alice and an external receiver which is named the eavesdropper (Eve). Accordingly, perfect knowledge of Eve’s channel CSI can not be assumed and hence information-theoretic security can not be introduced. This case is happened when Eve is unknown eavesdropper and that makes Alice unable to access it's CSI.  
\par In this paper, the main and wiretap channels are supposed to be independent and they subject to quasi-static FB fading channels. In addition, Alice, Bob, and Alice are equipped with a single antenna and perfect knowledge of the CSI of both Bob and Eve are assumed. When Alice transmits the signal $s(n)$, the received signals $r_l(n)$ at both Bob and Eve is given as [20]
\setcounter{equation}{0}
\label{eqn_1}
\begin{equation}
r_l(n)=h_l(n) s(n)+w_l(n).
\end{equation}
where $l \in \{D, E\}$, $D$, and $E$ stand for Bob, and eavesdropper, respectively. Moreover, $h_l(n)$ and $w_l(n)$ are the FB fading channel and the additive white Gaussian noise that has zero mean and fixed variance, respectively.

\subsection{The PDF and CDF of Fluctuating Beckmann Fading Channel Model}
$\textbf{Case\_1}$: The PDF of the instantaneous SNR $\gamma_l$, $f_{\gamma_l}(\gamma_l)$, for the destination (Bob), $D$, and the eavesdropper, $E$, channels using FB fading channel model is given by [22, eq. (5)]
\setcounter{equation}{1}
\label{eqn_2}
\begin{equation}
f_{\gamma_l}(\gamma_l)=\frac{\Omega_l}{\Gamma(\mu_l)}\gamma^{\mu_l-1}_l \Phi^{(4)}_2 \bigg(\frac{\mu_l}{2}-m_l,\frac{\mu_l}{2}-m_l,m_l,m_l;\mu_l;-\frac{\gamma_l}{\bar{\gamma}_l\sqrt{\eta_l \alpha_{2_l}}},-\frac{\gamma_l \sqrt{\eta_l}}{\bar{\gamma}_l\sqrt{\alpha_{2_l}}},-\frac{\gamma_l c_{1_l}}{\bar{\gamma}_l},-\frac{\gamma_l c_{2_l}}{\bar{\gamma}_l}\bigg).
\end{equation}
where $l \in \{D, E\}$, $\Omega_l=\frac{\alpha^{m_l-\frac{\mu_l}{2}}_{2l}}{\bar{\gamma}^{\mu_l}_l \alpha^{m_l}_{1l}}$, $\alpha_{2_l}=\frac{4 \eta_l}{\mu^2_l (1+\eta_l)^2(1+\kappa_l)^2}$, $\bar{\gamma}_l$ is the average SNR, $m_l$ is the shadowing severity parameter, $\Gamma(a)=\int_0^\infty x^{a-1} e^{-x} dx$ is the Gamma function and $\Phi^{(4)}_2(.)$ is the multivariate confluent hypergeometric function defined in [25, eq. (1.7.10)]. Furthermore, $c_{1l,2l}$ are the roots of $\alpha_{1l} s^2+ \beta_l s+1$ with
\setcounter{equation}{2}
\label{eqn_3}
\begin{align}
\alpha_{1_l}&=\alpha_{2_l}+\frac{2 \kappa_l (\varrho^2_l+\eta_l)}{m_l (1+\varrho^2_l) \mu_l (1+\eta_l) (1+\kappa_l)^2}. \nonumber\\
\beta_l&=-\frac{1}{1+\kappa_l}\bigg[\frac{2}{\mu_l}+\frac{\kappa_l}{m_l}\bigg].
\end{align}
where $\kappa_l=\frac{p^2_l+q^2_l}{\mu_l(\sigma^2_{x_l}+\sigma^2_{y_l})}$, $\varrho^2_l=\frac{p^2_l}{q^2_l}$, $\eta_l = \frac{\sigma^2_{x_l}}{\sigma^2_{y_l}}$, $\sigma^2_{x_l}=E[X^2_{i_l}]$, $\sigma^2_{y_l}=E[Y^2_{i_l}]$, $p_{i_l}$ and $q_{i_l}$ are real numbers for $i$th cluster and $X^2_{i_l}$ and $Y^2_{i_l}$ mutually independent Gaussian random processes.
\par The CDF of the FB fading channel condition is expressed as [22, eq. (6)] 
\setcounter{equation}{3}
\label{eqn_4}
\begin{equation}
F_{\gamma_l}(\gamma_l)=\frac{\Omega_l}{\Gamma(\mu_l+1)}\gamma^{\mu_l}_l \Phi^{(4)}_2 \bigg(\frac{\mu_l}{2}-m_l,\frac{\mu_l}{2}-m_l,m_l,m_l;\mu_l+1;-\frac{\gamma_l}{\bar{\gamma}_l\sqrt{\eta_l \alpha_{2_l}}},-\frac{\gamma_l \sqrt{\eta_l}}{\bar{\gamma}_l\sqrt{\alpha_{2_l}}},-\frac{\gamma_l c_{1_l}}{\bar{\gamma}_l},-\frac{\gamma_l c_{2_l}}{\bar{\gamma}_l}\bigg).
\end{equation}
\par $\textbf{Case\_2}$: When $m$ is integer number and $\mu$ is even number, the PDF and the CDF are, respectively, given by [22, eqs. (10) and (14)]
\setcounter{equation}{4}
\label{eqn_5}
\begin{equation}
f_{\gamma_l}(\gamma_l)=\Omega_l  \sum_{i_l=1}^{N_l(m_l,\mu_l)} e^{-\frac{\vartheta_l}{\bar{\gamma}_l}\gamma_l} \sum_{j_l=1}^{|\omega_{i_l}|}\frac{A_{{i_l}{j_l}}}{(j_l-1)!}\gamma^{j_l-1}_l.
\end{equation}
and
\label{eqn_6}
\begin{equation}
F_{\gamma_l}(\gamma_l)= 1+\Omega_l  \sum_{i_l=1}^{N_l(m_l,\mu_l)} e^{-\frac{\vartheta_l}{\bar{\gamma}_l}\gamma_l} \sum_{j_l=1}^{|\omega_{i_l}|}\frac{B_{{i_l}{j_l}}}{(j_l-1)!}\gamma^{j_l-1}_l
\end{equation}
where $\omega_l=[m_l, m_l, \frac{\mu_l}{2}-m_l, \frac{\mu_l}{2}-m_l]$, $\vartheta_l=[c_{1l}, c_{2l}, \frac{\mu_l (1+\eta_l)(1+\kappa_l)}{2 \eta_l}, \frac{\mu_l (1+\eta_l)(1+\kappa_l)}{2}]$, $N_l(m_l,\mu_l)= 2[1+\text{u}(\frac{\mu_l}{2}, m_l)]$, u(.) is the unit step function, and $A_{{i_l}{j_l}}$ and $B_{{i_l}{j_l}}$ are calculated by [22, eq.(51)] and [22, eq. (52)], respectively.
\section{Average Secrecy Capacity}
The normalised ASC that is defined as the difference between the capacity of the main and wiretap channels over instantaneous SNR, $\gamma$, can be calculated by $\bar{C}_s=I_1+I_2-I_3$ [15, eq. (6)] where $I_1$, $I_2$, and $I_3$ are respectively expressed as
\label{eqn_7}
\begin{equation}
I_1=\int_0^\infty \text{ln}(1+\gamma_D)f_D(\gamma_D)F_E(\gamma_D)d\gamma_D.
\end{equation} 

\label{eqn_8}
\begin{equation}
I_2=\int_0^\infty \text{ln}(1+\gamma_E)f_E(\gamma_E)F_D(\gamma_E)d\gamma_E.
\end{equation}

\label{eqn_9}
\begin{equation}
I_3=\int_0^\infty \text{ln}(1+\gamma_E)f_E(\gamma_E)d\gamma_E.
\end{equation}

\label{Theorem_1}
\begin{theorem}
The exact closed-form expressions for $I_1$, $I_2$, and $I_3$ using the PDF and the CDF of $\textbf{Case\_1}$ are given in (10), (11), and (12), respectively, at the bottom of this page where $H^{s,r:a,b;...;a_n,b_n}_{p,q:c,d;...,d_n,c_n}[.]$ denotes the EGBFHF that is defined in [26, A.1]. To the best of the authors knowledge, (10), (11), and (12) are novel. 
\end{theorem}

\label{Proof_Appendix_A}
\begin{IEEEproof}
See Appendix A.
\end{IEEEproof}
 
\label{eqn_10_11_12}
\setcounter{equation}{9}
\begin{table*}[h]
\hrulefill
\vspace*{1pt}
\begin{align}
I_1&=\frac{\Omega_D \Omega_E}{[\Gamma(\frac{\mu_D}{2}-m_D)\Gamma(\frac{\mu_E}{2}-m_E)\Gamma(m_D) \Gamma(m_E)]^2}  \nonumber\\ 
&\times H^{1,2:1,1;1,1;1,1;1,1;1,1;1,1;1,1;1,1}_{2,4:1,1;1,1;1,1;1,1;1,1;1,1;1,1;1,1} 
\bigg[ \begin{matrix}
  \frac{1}{\bar{\gamma}_D \sqrt{\eta_D \alpha_{2_D}}},
  \frac{\sqrt{\eta_D}}{\bar{\gamma}_D \sqrt{\alpha_{2_D}}},
  \frac{c_{1_D}}{\bar{\gamma}_D},
   \frac{c_{2_D}}{\bar{\gamma}_D},
   \frac{1}{\bar{\gamma}_E \sqrt{\eta_E \alpha_{2_E}}},
  \frac{\sqrt{\eta_E}}{\bar{\gamma}_E \sqrt{\alpha_{2_E}}},
  \frac{c_{1_E}}{\bar{\gamma}_E},
   \frac{c_{2_E}}{\bar{\gamma}_E}
\end{matrix} \bigg\vert  \nonumber\\
&\begin{matrix}
  (1-\mu_D-\mu_E;1,1,1,1,1,1,1,1), (1-\mu_D-\mu_E;1,1,1,1,1,1,1,1)\\
  (-\mu_D-\mu_E;1,1,1,1,1,1,1,1), (1-\mu_D;1,1,1,1,0,0,0,0), (-\mu_E;0,0,0,0,1,1,1,1)
\end{matrix}
\bigg\vert
\nonumber\\
&\begin{matrix}
  (1+m_D-\frac{\mu_D}{2},1)\\
  (0,1)
\end{matrix} 
\bigg\vert  
\begin{matrix}
  (1+m_D-\frac{\mu_D}{2},1)\\
  (0,1)
\end{matrix} 
\bigg\vert
\begin{matrix}
  (1-m_D,1)\\
  (0,1)
\end{matrix} 
\bigg\vert
\begin{matrix}
  (1-m_D,1)\\
  (0,1)
\end{matrix} 
\bigg\vert\nonumber\\
&\begin{matrix}
  (1+m_E-\frac{\mu_E}{2},1)\\
  (0,1)
\end{matrix} 
\bigg\vert 
\begin{matrix}
  (1+m_E-\frac{\mu_E}{2},1)\\
  (0,1)
\end{matrix} 
\bigg\vert 
\begin{matrix}
  (1-m_E,1)\\
  (0,1)
\end{matrix} 
\bigg\vert
\begin{matrix}
  (1-m_E,1)\\
  (0,1)
\end{matrix}
\bigg]
\end{align}
\hrulefill
\vspace*{1pt}
\label{eqn_11}
\begin{align}
I_2&=\frac{\Omega_E \Omega_D}{[\Gamma(\frac{\mu_E}{2}-m_E)\Gamma(\frac{\mu_D}{2}-m_D)\Gamma(m_E) \Gamma(m_D)]^2} \nonumber\\  
&\times H^{1,2:1,1;1,1;1,1;1,1;1,1;1,1;1,1;1,1}_{2,4:1,1;1,1;1,1;1,1;1,1;1,1;1,1;1,1} 
\bigg[ \begin{matrix}
  \frac{1}{\bar{\gamma}_E \sqrt{\eta_E \alpha_{2_E}}},
  \frac{\sqrt{\eta_E}}{\bar{\gamma}_E \sqrt{\alpha_{2_E}}},
  \frac{c_{1_E}}{\bar{\gamma}_E},
   \frac{c_{2_E}}{\bar{\gamma}_E},
   \frac{1}{\bar{\gamma}_D \sqrt{\eta_D \alpha_{2_D}}},
  \frac{\sqrt{\eta_D}}{\bar{\gamma}_D \sqrt{\alpha_{2_D}}},
  \frac{c_{1_D}}{\bar{\gamma}_D},
   \frac{c_{2_D}}{\bar{\gamma}_D}
\end{matrix} \bigg\vert  \nonumber\\
&\begin{matrix}
  (1-\mu_E-\mu_D;1,1,1,1,1,1,1,1), (1-\mu_E-\mu_D;1,1,1,1,1,1,1,1)\\
  (-\mu_E-\mu_D;1,1,1,1,1,1,1,1), (1-\mu_E;1,1,1,1,0,0,0,0), (-\mu_D;0,0,0,0,1,1,1,1)\\
\end{matrix}
\bigg\vert\nonumber\\
&\begin{matrix}
  (1+m_E-\frac{\mu_E}{2},1)\\
  (0,1)
\end{matrix} 
\bigg\vert  
\begin{matrix}
  (1+m_E-\frac{\mu_E}{2},1)\\
  (0,1)
\end{matrix} 
\bigg\vert
\begin{matrix}
  (1-m_E,1)\\
  (0,1)
\end{matrix} 
\bigg\vert
\begin{matrix}
  (1-m_E,1)\\
  (0,1)
\end{matrix} 
\bigg\vert\nonumber\\
&\begin{matrix}
  (1+m_D-\frac{\mu_D}{2},1)\\
  (0,1)
\end{matrix} 
\bigg\vert 
\begin{matrix}
  (1+m_D-\frac{\mu_D}{2},1)\\
  (0,1)
\end{matrix} 
\bigg\vert 
\begin{matrix}
  (1-m_D,1)\\
  (0,1)
\end{matrix} 
\bigg\vert
\begin{matrix}
  (1-m_D,1)\\
  (0,1)
\end{matrix}
\bigg]
\end{align}
\hrulefill
\vspace*{1pt}
\label{eqn_12}
\begin{align}
I_3&=\frac{ \Omega_E }{[\Gamma(\mu_E-m_E) \Gamma(m_E)]^2}H^{1,2:1,1;1,1;1,1;1,1}_{2,3:1,1;1,1;1,1;1,1} 
\begin{matrix}
\bigg[\frac{1}{\bar{\gamma}_E \sqrt{\eta_E \alpha_{2_E}}},
  \frac{\sqrt{\eta_E}}{\bar{\gamma}_E \sqrt{\alpha_{2_E}}},
  \frac{c_{1_E}}{\bar{\gamma}_E},
   \frac{c_{2_E}}{\bar{\gamma}_E} 
   \end{matrix}\bigg\vert \nonumber\\
&\begin{matrix}
  (1-\mu_E,1),(1-\mu_E,1)\\
  (1-\mu_E,1),(-\mu_E,1),(-\mu_E,1)
\end{matrix} \bigg\vert 
\begin{matrix}
  (1+m_E-\frac{\mu_E}{2},1)\\
  (0,1)
\end{matrix} \bigg\vert
\begin{matrix}
  (1+m_E-\frac{\mu_E}{2},1)\\
  (0,1)
\end{matrix} \bigg\vert 
 \begin{matrix}
  (1-m_E,1)\\
  (0,1)
\end{matrix}\bigg\vert 
 \begin{matrix}
  (1-m_E,1)\\
  (0,1)
\end{matrix}
\bigg]
\end{align} 
\end{table*}

\label{corollary_1}
\begin{corollary}
For \textbf{Case 2}, $I_1$, $I_2$, and $I_3$ are respectively derived in simple analytic exact closed-form expressions as shown in (13), (14), and (15), at the top this page. Substituting these expressions in $\bar{C}_s=I_1+I_2-I_3$ and performing some straightforward manipulations, novel result is obtained as given in (16) at the top of this page.
\end{corollary}
\label{Proof_Appendix_B}
\begin{IEEEproof}
See Appendix B.
\end{IEEEproof}

\label{eqn_13_14_15_16}
\begin{table*}
\begin{align}
I_1&=\Omega_D \sum_{i_D=1}^{N_D (m_D,\mu_D)} \sum_{j_D=1}^{|\omega_{i_D}|}\frac{A_{{i_D}{j_D}}}{(j_D-1)!} \Bigg[\Gamma(j_D) e^{\frac{\vartheta_{i_D}}{\bar{\gamma}_D}} \sum_{k=1}^{j_D} \frac{\Gamma \vast(k-j_D, \frac{\vartheta_{i_D}}{\bar{\gamma}_D}\vast)}{\vast(\frac{\vartheta_{i_D}}{\bar{\gamma}_D}\vast)^k}+ \nonumber \\
&\Omega_E \sum_{i_E=1}^{N_E (m_E,\mu_E)} \sum_{j_E=1}^{|\omega_{i_E}|} 
\frac{B_{{i_E}{j_E}}}{(j_E-1)!} \Gamma(j_D+j_E-1) e^{\big(\frac{\vartheta_{i_D}}{\bar{\gamma}_D}+\frac{\vartheta_{i_E}}{\bar{\gamma}_E}\big)} \sum_{r=1}^{j_D+j_E-1} \frac{\Gamma \vast(r-j_D-j_E+1,\frac{\vartheta_{i_D}}{\bar{\gamma}_D}+\frac{\vartheta_{i_E}}{\bar{\gamma}_E} \vast)}{\vast(\frac{\vartheta_{i_D}}{\bar{\gamma}_D}+\frac{\vartheta_{i_E}}{\bar{\gamma}_E}\vast)^{r-j_D-j_E+1}}\Bigg].
\end{align}
\hrulefill
\vspace*{1pt}

\label{eqn_14}
\begin{align}
I_2&=\Omega_E \sum_{i_E=1}^{N_E (m_E,\mu_E)} \sum_{j_E=1}^{|\omega_{i_E}|}\frac{A_{{i_E}{j_E}}}{(j_E-1)!} \bigg[\Gamma(j_E) e^{\frac{\vartheta_{i_E}}{\bar{\gamma}_E}} \sum_{k=1}^{j_E} \frac{\Gamma \vast(k-j_E, \frac{\vartheta_{i_E}}{\bar{\gamma}_E}\vast)}{\vast(\frac{\vartheta_{i_E}}{\bar{\gamma}_E}\vast)^k}+ \nonumber \\
&\Omega_D \sum_{i_D=1}^{N_D (m_D,\mu_D)} \sum_{j_D=1}^{|\omega_{i_D}|} 
\frac{B_{{i_D}{j_D}}}{(j_D-1)!} \Gamma(j_E+j_D-1) e^{\big(\frac{\vartheta_{i_E}}{\bar{\gamma}_E}+\frac{\vartheta_{i_D}}{\bar{\gamma}_D}\big)} \sum_{r=1}^{j_E+j_D-1} \frac{\Gamma \vast(r-j_E-j_D+1,\frac{\vartheta_{i_E}}{\bar{\gamma}_E}+\frac{\vartheta_{i_D}}{\bar{\gamma}_D} \vast)}{\vast(\frac{\vartheta_{i_E}}{\bar{\gamma}_E}+\frac{\vartheta_{i_D}}{\bar{\gamma}_D}\vast)^{r-j_E-j_D+1}}\bigg]
\end{align}
\hrulefill
\vspace*{1pt}
\label{eqn_15}
\begin{align}
I_3=\Omega_E \sum_{i_E=1}^{N_E (m_E,\mu_E)} \sum_{j_E=1}^{|\omega_{i_E}|} A_{{i_E}{j_E}} e^{\frac{\vartheta_{i_E}}{\bar{\gamma}_E}} \sum_{k=1}^{j_E} \frac{\Gamma \vast(k-j_E, \frac{\vartheta_{i_E}}{\bar{\gamma}_E}\vast)}{\vast(\frac{\vartheta_{i_E}}{\bar{\gamma}_E}\vast)^k}
\end{align}
\hrulefill
\vspace*{1pt}
\label{eqn_16}
\begin{align}
\bar{C}_s=& \Omega_D \sum_{i_D=1}^{N_D (m_D,\mu_D)} \sum_{j_D=1}^{|\omega_{i_D}|} A_{{i_D}{j_D}} e^{\frac{\vartheta_{i_D}}{\bar{\gamma}_D}} \sum_{k=1}^{j_D} \frac{\Gamma \vast(k-j_D, \frac{\vartheta_{i_D}}{\bar{\gamma}_D}\vast)}{\vast(\frac{\vartheta_{i_D}}{\bar{\gamma}_D}\vast)^k} \nonumber \\
& +\Omega_D \Omega_E \sum_{i_D=1}^{N_D (m_D,\mu_D)} \sum_{j_D=1}^{|\omega_{i_D}|} \sum_{i_E=1}^{N_E (m_E,\mu_E)} \sum_{j_E=1}^{|\omega_{i_E}|} 
 \frac{\Gamma(j_D+j_E-1)}{\Gamma(j_E) \Gamma(j_D)}   [A_{{i_D}{j_D}} B_{{i_E}{j_E}}+A_{{i_E}{j_E}} B_{{i_D}{j_D}}] e^{\big(\frac{\vartheta_{i_D}}{\bar{\gamma}_D}+\frac{\vartheta_{i_E}}{\bar{\gamma}_E}\big)} \nonumber \\ 
& \sum_{r=1}^{j_D+j_E-1} \frac{\Gamma \vast(r-j_D-j_E+1,\frac{\vartheta_{i_D}}{\bar{\gamma}_D}+\frac{\vartheta_{i_E}}{\bar{\gamma}_E} \vast)}{\vast(\frac{\vartheta_{i_D}}{\bar{\gamma}_D}+\frac{\vartheta_{i_E}}{\bar{\gamma}_E}\vast)^{r-j_D-j_E+1}} .
\end{align}
\hrulefill
\vspace*{1pt}  
 \end{table*}
 
\section{Secure Outage Probability}
The SOP is defined as the probability of falling the instantaneous secrecy capacity, $C_s$, of the system below the target secrecy threshold, $R_s$, i.e., $\mathcal{P}(C_s < R_s)$ where $\mathcal{P}(.)$ stands for the probability symbol. Mathematically, the SOP can be evaluated by [20, eq. (20)]
\label{eqn_17}
\begin{equation}
\text{SOP}=\int_0^\infty F_D(\theta \gamma_E+\theta-1)f_E(\gamma_E)d\gamma_E
\end{equation}
where $\theta = \mathrm{exp}(R_s) \geq 1$ with $R_s \geq 0$ denotes the target secrecy threshold. 

\label{Theorem_2}
\begin{theorem}
The SOP for $\textbf{Case\_1}$ and $\textbf{Case\_2}$ of the PDF and the CDF are expressed in exact closed-form as shown in (18) and (19), respectively. In (19), ${{b}\choose{a}}\triangleq \frac{b!}{(b-a)!}$ stands for the binomial coefficient [25, eq. (1.1.16)] and ($a$)$_r$ is the Pochhammer symbol [25, eq. (1.1.15)]. Additionally, (19) is simpler than (18) and hence better insights can be obtained for the SOP. To the best authors' knowledge, (18) and (19) are novel and mathematically tractable. 
\end{theorem}
\label{Proof_Appendix_C}
\begin{IEEEproof}
See Appendix C.
\end{IEEEproof}

\label{eqn_18}
\begin{table*}[h]
\hrulefill
\vspace*{1pt}  
\begin{align}
\text{SOP}&=\frac{\Omega_E \Omega_D (\theta-1)^{\mu_D+\mu_E}}{\theta^{\mu_E}[\Gamma(\frac{\mu_E}{2}-m_E)\Gamma(\frac{\mu_D}{2}-m_D)\Gamma(m_E) \Gamma(m_D)]^2}  \nonumber\\ 
&\times H^{1,1:1,1;1,1;1,1;1,1;1,1;1,1;1,1;1,1}_{2,3:1,1;1,1;1,1;1,1;1,1;1,1;1,1;1,1} 
\bigg[ \begin{matrix}
  \frac{\theta-1}{\theta \bar{\gamma}_E \sqrt{\eta_E \alpha_{2_E}}},
  \frac{(\theta-1)\sqrt{\eta_E}}{\theta \bar{\gamma}_E \sqrt{\alpha_{2_E}}},
  \frac{(\theta-1) c_{1_E}}{\theta \bar{\gamma}_E},
   \frac{(\theta-1) c_{2_E}}{\theta \bar{\gamma}_E},
   \frac{\theta-1}{\bar{\gamma}_D \sqrt{\eta_D \alpha_{2_D}}},
  \frac{(\theta-1) \sqrt{\eta_D}}{\bar{\gamma}_D \sqrt{\alpha_{2_D}}},
\end{matrix} \nonumber\\
&\begin{matrix}
\frac{(\theta-1) c_{1_D}}{\bar{\gamma}_D},\frac{(\theta-1) c_{2_D}}{\bar{\gamma}_D}
\end{matrix} \bigg\vert  
\begin{matrix}
  (1-\mu_E;1,1,1,1,0,0,0,0), (-\mu_D;0,0,0,0,1,1,1,1)\\
  (-\mu_E-\mu_D;1,1,1,1,1,1,1,1), (-\mu_E;1,1,1,1,0,0,0,0), (-\mu_D;0,0,0,0,1,1,1,1)
\end{matrix}
\bigg\vert
\nonumber\\
&\begin{matrix}
  (1+m_E-\frac{\mu_E}{2},1)\\
  (0,1)
\end{matrix} 
\bigg\vert  
\begin{matrix}
  (1+m_E-\frac{\mu_E}{2},1)\\
  (0,1)
\end{matrix} 
\bigg\vert
\begin{matrix}
  (1-m_E,1)\\
  (0,1)
\end{matrix} 
\bigg\vert
\begin{matrix}
  (1-m_E,1)\\
  (0,1)
\end{matrix} 
\bigg\vert
\begin{matrix}
  (1+m_D-\frac{\mu_D}{2},1)\\
  (0,1)
\end{matrix} 
\bigg\vert \nonumber\\
&\begin{matrix}
  (1+m_D-\frac{\mu_D}{2},1)\\
  (0,1)
\end{matrix} 
\bigg\vert
\begin{matrix}
  (1-m_D,1)\\
  (0,1)
\end{matrix} 
\bigg\vert
\begin{matrix}
  (1-m_D,1)\\
  (0,1)
\end{matrix}
\bigg]
\end{align}
\hrulefill
\vspace*{1pt} 
\label{eqn_19}
\begin{align}
\text{SOP}=1+\Omega_D \Omega_E  \sum_{i_E=1}^{N_E (m_E,\mu_E)} \sum_{j_E=1}^{|\omega_{i_E}|} A_{{i_E}{j_E}} \sum_{i_D=1}^{N_D (m_D,\mu_D)} \sum_{j_D=1}^{|\omega_{i_D}|} \frac{B_{{i_D}{j_D}}}{\Gamma(j_D)}
e^{\frac{\vartheta_{i_D}}{\bar{\gamma}_D}(1-\theta)} \sum_{r=0}^{j_D-1}   {{j_D-1}\choose{r}} \nonumber\\
\frac{\theta^r (j_E)_r }{(\theta-1)^{r-j_D+1} \vast(\frac{\theta\vartheta_{i_D}}{\bar{\gamma}_D}+\frac{\vartheta_{i_E}}{\bar{\gamma}_E}\vast)^{r+j_E}}
\end{align}
\hrulefill
\vspace*{1pt}
\end{table*}

\section{Lower Bound of SoP}
According to [6], the SOP$^L$ can be obtained from (17) when $\gamma_E$ tends to $\infty$. Consequently, the SOP$^L$ can be computed by
\label{eqn_20}
\begin{align}
\text{SOP}^L &=\int_0^\infty F_D(\theta \gamma_E) f_E(\gamma_E) d\gamma_E \nonumber\\
& \leq \text{SOP}
\end{align}

\label{Theorem_3}
\begin{theorem}
The $\text{SOP}^L$ when $\mu_l$ and $m_l$ are arbitrary numbers, namely, $\textbf{Case\_1}$, is presented in (21) whereas for $\textbf{Case\_2}$, the $\text{SOP}^L$ is given in (22) where $\phi=\frac{\theta}{\bar{\gamma}_D \sqrt{\eta_D \alpha_{2_D}}}+\frac{1}{\bar{\gamma}_E \sqrt{\eta_E \alpha_{2_E}}}$. One can see that (21) and (22) are obtained in simple exact closed-form expressions. To the best of our knowledge, (21) and (22) are also new.
\end{theorem}
\label{App_D}
\begin{IEEEproof}
See Appendix D.
\end{IEEEproof}

\label{eqn_21}
\begin{table*}[h]
\hrulefill
\vspace*{1pt}
\begin{align}
\text{SOP}^L&=\frac{\Omega_E \Omega_D \theta^{\mu_D}}{\phi^{\mu_E+\mu_D}\Gamma(\frac{\mu_E}{2}-m_E)\Gamma(\frac{\mu_D}{2}-m_D)[\Gamma(m_E) \Gamma(m_D)]^2} \nonumber\\ 
&\times H^{0,1:1,1;1,1;1,1;1,1;1,1;1,1}_{1,2:1,1;1,1;1,1;1,1;1,1;1,1} 
\bigg[\begin{matrix}
  \frac{\eta_E-1}{\phi \bar{\gamma}_E \sqrt{\eta_E \alpha_{2_E}}},
  \frac{\sqrt{\eta_E \alpha_{2_E}} c_{1_E}-1}{\phi \bar{\gamma}_E \sqrt{\eta_E \alpha_{2_E}}},
  \frac{\sqrt{\eta_E \alpha_{2_E}} c_{2_E}-1}{\phi \bar{\gamma}_E \sqrt{\eta_E \alpha_{2_E}}},
   \frac{(\eta_D-1) \theta}{\phi \bar{\gamma}_D \sqrt{\eta_D \alpha_{2_D}}},
     \frac{(\sqrt{\eta_D \alpha_{2_D}} c_{1_D}-1)\theta}{\phi \bar{\gamma}_D \sqrt{\eta_D \alpha_{2_D}}},
   \end{matrix}
   \nonumber\\
  & \begin{matrix}
   \frac{(\sqrt{\eta_D \alpha_{2_D}} c_{2_D}-1)\theta}{\phi \bar{\gamma}_D \sqrt{\eta_D \alpha_{2_D}}}
\bigg\vert 
\end{matrix}
\begin{matrix}
  (1-\mu_D-\mu_E;1,1,1,1,1,1)\\
  (1-\mu_E;1,1,1,0,0,0),(-\mu_D;0,0,0,1,1,1)
\end{matrix}
\bigg\vert
\begin{matrix}
  (1+m_E-\frac{\mu_E}{2},1)\\
  (0,1)
\end{matrix} 
\bigg\vert
\begin{matrix}
  (1-m_E,1)\\
  (0,1)
\end{matrix} 
\bigg\vert 
 \nonumber\\
& \begin{matrix}
  (1-m_E,1)\\
  (0,1)
\end{matrix} 
\bigg\vert 
\begin{matrix}
  (1+m_D-\frac{\mu_D}{2},1)\\
  (0,1)
\end{matrix} 
\bigg\vert
\begin{matrix}
  (1-m_D,1)\\
  (0,1)
\end{matrix} 
\bigg\vert
\begin{matrix}
  (1-m_D,1)\\
  (0,1)
\end{matrix}
\bigg]
\end{align}
\hrulefill
\vspace*{1pt}
\label{eqn_22}
\begin{align}
\text{SOP}^L&=1+\Omega_E \Omega_D  \sum_{i_E=1}^{N_E (m_E,\mu_E)} \sum_{j_E=1}^{|\omega_{i_E}|} \frac{A_{{i_E}{j_E}}}{\Gamma(j_E)} \sum_{i_D=1}^{N_D (m_D,\mu_D)} \sum_{j_D=1}^{|\omega_{i_D}|} \theta^{j_D-1}\frac{ \Gamma(j_D+j_E-1)}{\Gamma(j_D)}
 \frac{B_{{i_D}{j_D}}}{\vast(\frac{\theta\vartheta_{i_D}}{\bar{\gamma}_D}+\frac{\vartheta_{i_E}}{\bar{\gamma}_E}\vast)^{j_D+j_E-1}}
\end{align}
\hrulefill
\vspace*{1pt}
\end{table*}

\section{Probability of Strictly Positive Secrecy Capacity}
The SPSC refers to the probability of positive $C_s$, namely, $\mathcal{P}(C_s>0)$. Therefore, it can be calculated by [18, eq. (11)]
\label{eqn_23}
\begin{equation}
\text{SPSC}=1-\text{SOP}^L \quad \text{for} \quad \theta=1 
\end{equation}
\par It can be observed that the SPSC for $\textbf{Case\_1}$ and $\textbf{Case\_2}$ can be obtained from (21) and (22), respectively, via substituting $\theta = 1$ and plugging the results in (23).

\section{Physical Layer Security over Special Cases of FB Fading Channel Model}
It has been mentioned in [22], the FB fading channel is a versatile model for nearly most of the well known distributions. Accordingly, our derived secrecy performance metrics can be utilised for different scenarios of wireless communications channels that have not been yet done in the literature due to the complexity of their statistical characterization. For example, when $\kappa = K$, $\mu = 1$, $m \rightarrow \infty$, $\eta = q$, and $\varrho = r$ the Beckmann distribution is the result. Furthermore, the PDF and the CDF of $\kappa-mu$ shadowed fading obtained from (2) or (5) and (4) or (6), respectively after plugging plugging $\kappa = K$, $\mu = \mu$, $m = m$, $\eta = 1$, and $\forall \varrho $. The ASC, the SOP, the SOP$^L$, and the SPSC for $\eta-mu$ fading condition can be deduced by substituting $\kappa = 0$, $\mu = \mu$, $\eta = \eta$ and vanishing of both $m$ and $\varrho$ in our derived expressions.
\label{fig_1}
\begin{figure}[h!]
\centering
  \includegraphics[width=3.8 in, height=2.74 in]{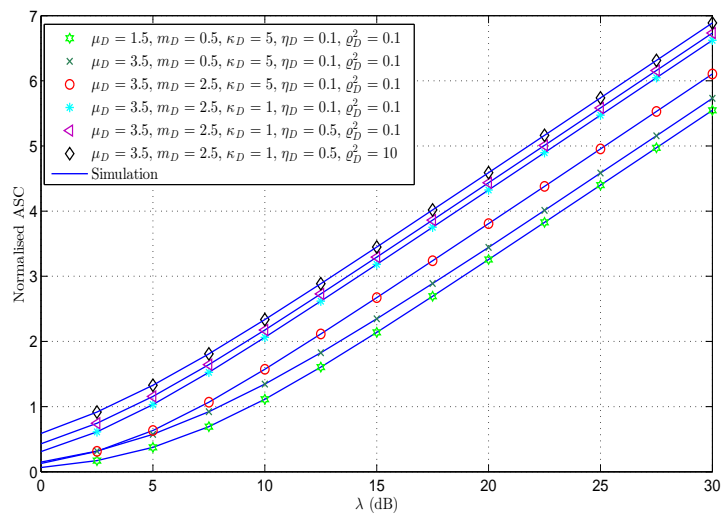} 
\caption{ASC versus $\lambda$ for $\bar{\gamma}_E = 5$ dB, $\mu_E = 1.5$, $m_E = 1.5$, $\kappa_E = 1$, $\eta_E = 0.1$, $\varrho^2_E = 0.1$ and different values of the fading parameters of Bob.}
\end{figure}
\label{fig_3}
\begin{figure}[h!]
\centering
  \includegraphics[width=3.8 in, height=2.74 in]{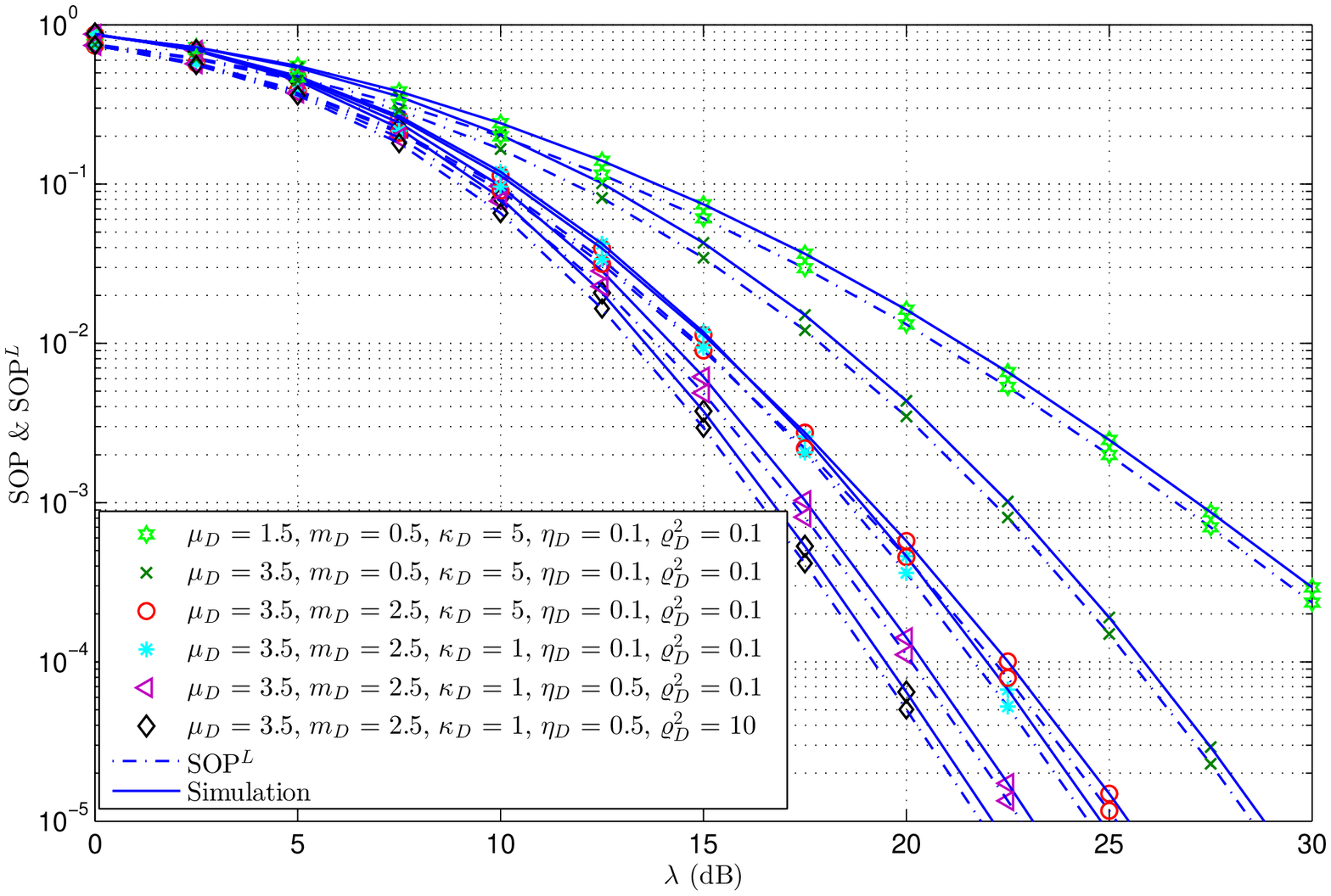} 
\caption{SOP and SOP$^L$ versus $\lambda$ for $\bar{\gamma}_E = 5$ dB, $\mu_E = 1.5$, $m_E = 1.5$, $\kappa_E = 1$, $\eta_E = 0.1$, $\varrho^2_E = 0.1$, $R_s = 1$ and different values of the fading parameters of Bob.}
\end{figure}
\label{fig_5}
  \begin{figure}[h!]
\centering
  \includegraphics[width=3.8 in, height=2.74 in]{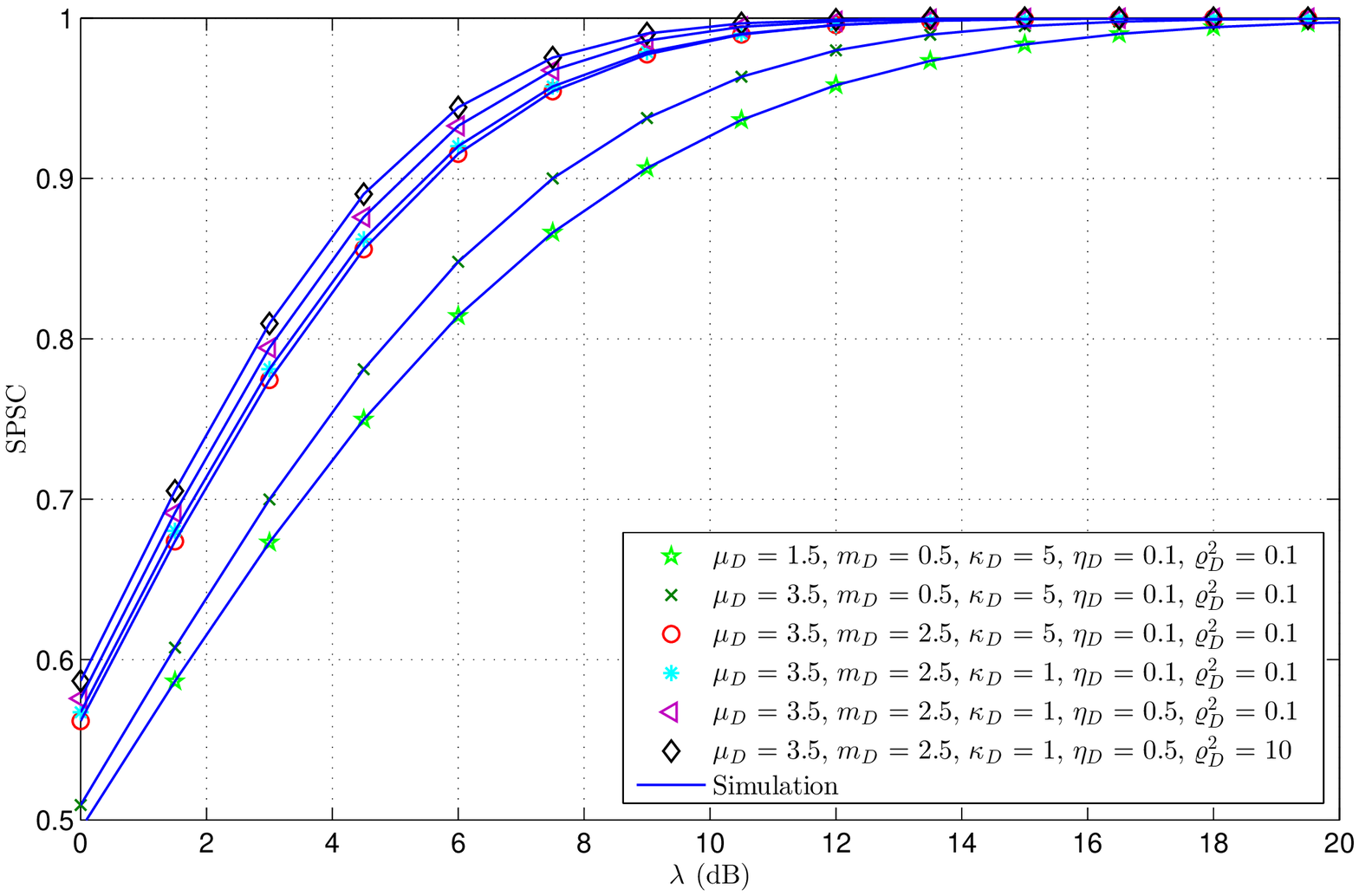} 
\caption{SPSC versus $\lambda$ for $\bar{\gamma}_E = 5$ dB, $\mu_E = 1.5$, $m_E = 1.5$, $\kappa_E = 1$, $\eta_E = 0.1$, $\varrho^2_E = 0.1$, $R_s = 1$ and different values of the fading parameters of Bob.}
\end{figure}

\label{fig_2}
\begin{figure}[h!]
\centering
  \includegraphics[width=3.8 in, height=2.74 in]{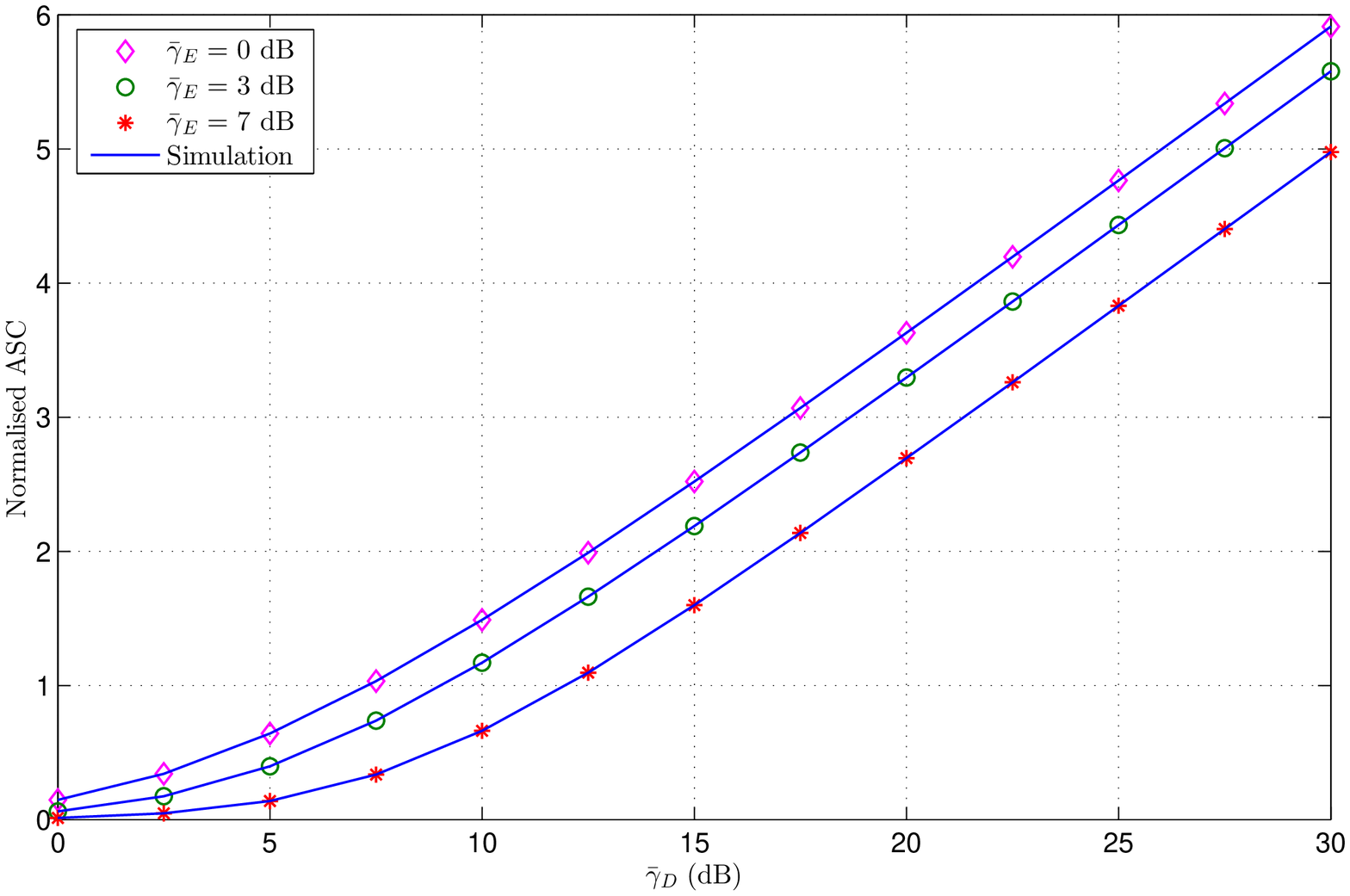} 
\caption{ASC versus $\bar{\gamma}_D$ for $\mu_D = \mu_E = 2.5$, $m_D = m_E = 1.5$, $\kappa_D = \kappa_E = 3$, $\eta_D = \eta_E = 0.5$, $\varrho^2_D = \varrho^2_E = 0.2$ and different values of $\bar{\gamma}_E$.}
\end{figure}
\label{fig_4}
\begin{figure}[h!]
\centering
  \includegraphics[width=3.8 in, height=2.74 in]{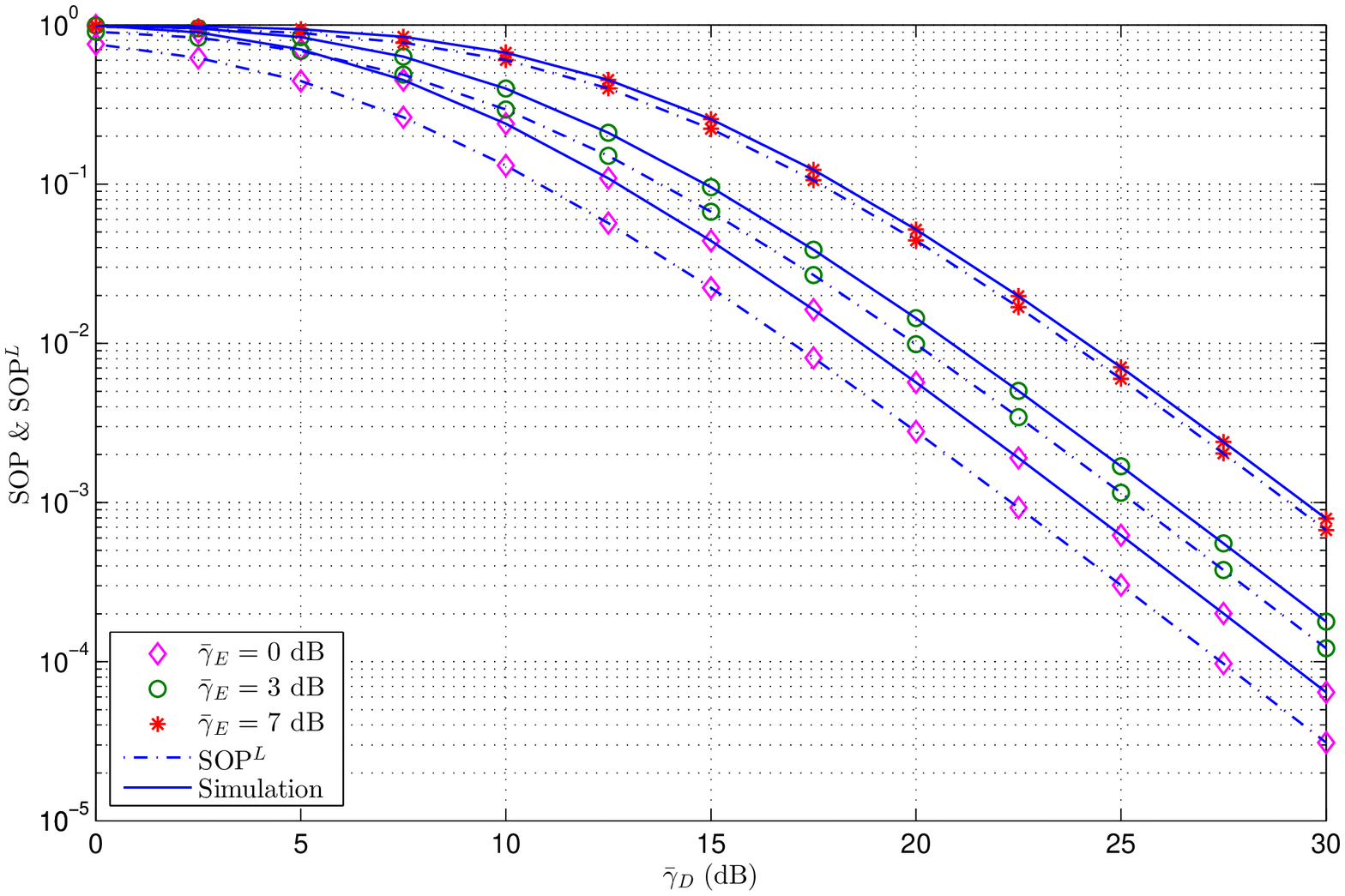} 
 \caption{SOP $\&$ SOP$^L$ versus $\bar{\gamma}_D$ for $\mu_D = \mu_E = 2.5$, $m_D = m_E = 1.5$, $\kappa_D = \kappa_E = 3$, $\eta_D = \eta_E = 0.5$, $\varrho^2_D = \varrho^2_E = 0.2$, $R_s = 1$ and different values of $\bar{\gamma}_E$.}
 \end{figure} 
 \label{fig_6}
  \begin{figure}[h!]
\centering
  \includegraphics[width=3.8 in, height=2.74 in]{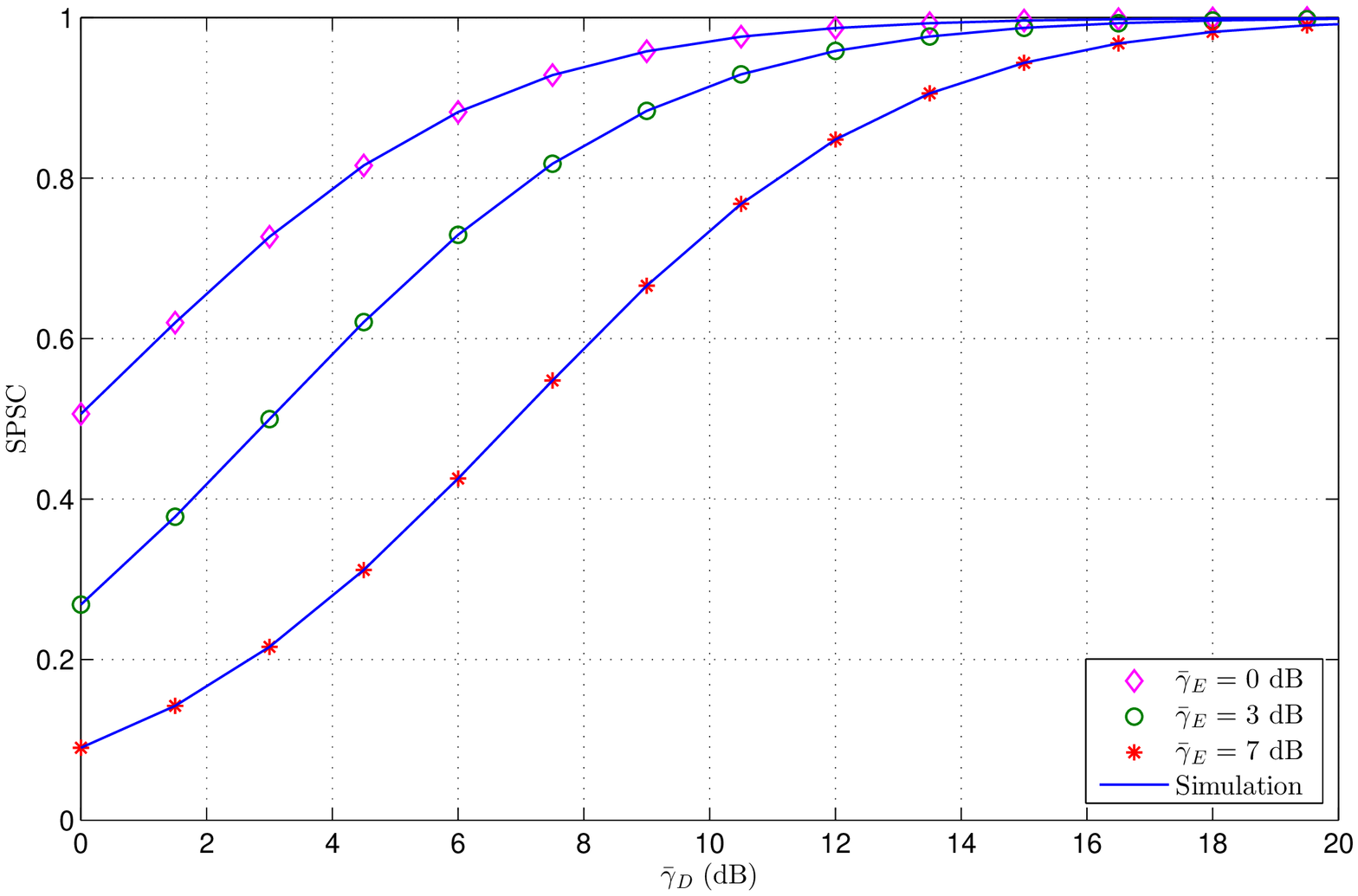}   
 \caption{SPSC versus $\bar{\gamma}_D$ for $\mu_D = \mu_E = 2.5$, $m_D = m_E = 1.5$, $\kappa_D = \kappa_E = 3$, $\eta_D = \eta_E = 0.5$, $\varrho^2_D = \varrho^2_E = 0.2$, $R_s = 1$ and different values of $\bar{\gamma}_E$.}
 \end{figure}
  
\section{Numerical and Simulation Results}
In this section, the numerical results of this work are verified via Monte Carlo simulations with $10^7$ realizations. The parameters of main and wiretap channels are assumed to be independent and non-identically distributed random variables. In all figures, the markers represents the numerical results, whereas the solid lines explain their simulation counterparts.
\par Figs. 1, 2 and 3 illustrate the ASC, the SOP $\&$ SOP$^L$ and the SPSC versus $\lambda = \bar{\gamma}_D/\bar{\gamma}_E$, respectively, for $\bar{\gamma}_E = 5$ dB, $\mu_E = 1.5$, $m_E = 1.5$, $\kappa_E = 1$, $\eta_E = 0.1$, $\varrho^2_E = 0.1$ and different values of the fading parameters of Bob. From these figures, one can see that the secrecy performance improves when $\mu_D$, $m_D$, $\eta_D$, or/and $\varrho^2_D$ increase. This is because the increasing in $\mu_D$ or/and $m_D$ correspond to a large number of the multipath clusters and the less shadowing impact at the Bob, respectively. Moreover, the improving in the values of $\eta_D$, or/and $\varrho^2_D$ mean a high power rate at the Bob, i.e., the total power of the in-phase components. on the contrary, the decreasing of $\kappa_D$ increases the values of all the studied performance metrics. The main reason is the parameter $\kappa_D$ represents the ratio between the total power of the dominant components and the total power of the scattered waves.
For example, in Fig. 1, at $m_D = 0.5$, $\kappa_D = 5$, $\eta_D = 0.1$, $\varrho^2 = 0.1$ and $\lambda = 15$ dB (fixed), the ASC for $\mu_D = 3.5$ is nearly $10\%$ higher than $\mu_D = 1.5$. In the same context, when $\mu_D = 3.5$ and $m_D$ changes from $0.5$ to $2.5$, the ASC is increased by roughly $25\%$. Furthermore, the ASC for the scenario $\mu_D = 3.5$, $m_D=2.5$, $\kappa_D = 1$, $\eta_D =0.1$, and $\varrho^2_D = 0.1$ at $\lambda = 15$ dB is approximately $2.139$ whereas for $\eta_D =0.5$ is $3.293$ which is less than for $\varrho^2_D = 10$ by nearly $5\%$. On the other side, the ASC for the previous scenario is decreased by roughly $16\%$ when $\kappa_D$ becomes 5. The impacts of the fading parameters on the provided results in Fig. 1 are confirmed by Figs. 2 and 3. In addition, from these figures, it is clear that the secrecy performance enhances when $\lambda$ increases. This refers to the high $\bar{\gamma}_D$ in comparison with the $\bar{\gamma}_E$ which would lead to make the Alice-Bob channel better than the Alice-Eve channel. 
\par Figs. 4, 5, and 6 demonstrate the ASC, the SOP $\&$ SOP$^L$ and the SPSC versus $\bar{\gamma}_D$, respectively, for $\mu_D = \mu_E = 2.5$, $m_D = m_E = 1.5$, $\kappa_D = \kappa_E = 3$, $\eta_D = \eta_E = 0.5$, $\varrho^2_D = \varrho^2_E = 0.2$ and different values of $\bar{\gamma}_E$. As expected, a clear improvement can be noticed in the secrecy performance of the considered system when $\bar{\gamma}_E$ reduces. The reason is the large deterioration of the wiretap channel. For instance, in Fig. 4, at $\bar{\gamma}_D = 15$ dB (fixed), the ASC for $\bar{\gamma}_E = 7$ dB and $\bar{\gamma}_E = 3$ dB are approximately $1.598$ and $2.189$, respectively.    
\par Another confirmation is presented in Figs. 2 and 5 via providing both SOP and SOP$^L$ which is less than or equal to the SOP of the same scenario.    
\par More importantly, the numerical results and Monte Carlo simulations are in perfect match for any given scenario.

\section{Conclusions}
\par This paper was dedicated to study the secrecy behaviour of the physical layer over Fluctuating Beckmann fading channel model. To be specific, the ASC, the SOP, the SOP$^L$, and the SPSC, were derived in novel exact closed-form expressions by using two cases for the values of the fading parameters. In the first case, the derived results were expressed in terms of EGBFHF. On the other side, the second case provided simple exact mathematically tractable closed-form expressions via assuming $\mu$ and $m$ for both Bob and Eve are even number and integer number, respectively. From the given results, a reduction in the values of the ASC, the SOP, the SOP$^L$, and the SPSC can be observed when the value of $\mu$, $m$, $\eta$, or/and $\varrho^2$ of the Bob decrease. However, the secrecy performances of the system improves when $\kappa$ of the Bob or/and $\gamma_E$ reduce. The results of this work can be employed to analyse the secrecy performance of the physical layer over a variety of fading channels with simple exact closed-form expressions.

\section*{Appendix A}
\section*{Proof of Theorem 1}
Substituting (2) and (4) in (7), this yields
\label{eqn_24}
\begin{align}
I_1&=\frac{\Omega_D \Omega_E}{\Gamma(\mu_D) \Gamma(\mu_E+1)}\int_0^\infty \text{ln}(1+\gamma_D) \gamma^{\mu_D+\mu_E-1}_D \nonumber\\
&\times \Phi^{(4)}_2 \bigg(\frac{\mu_D}{2}-m_D,\frac{\mu_D}{2}-m_D,m_D,m_D;\mu_D;-\frac{\gamma_D}{\bar{\gamma}_D\sqrt{\eta_D \alpha_{2_D}}},-\frac{\gamma_D \sqrt{\eta_D}}{\bar{\gamma}_D\sqrt{\alpha_{2_D}}},-\frac{\gamma_D c_{1_D}}{\bar{\gamma}_D},-\frac{\gamma_D c_{2_D}}{\bar{\gamma}_D}\bigg) \nonumber\\
&\times \Phi^{(4)}_2 \bigg(\frac{\mu_E}{2}-m_E,\frac{\mu_E}{2}-m_E,m_E,m_E;\mu_E+1;-\frac{\gamma_D}{\bar{\gamma}_E\sqrt{\eta_E \alpha_{2_E}}},-\frac{\gamma_D \sqrt{\eta_E}}{\bar{\gamma}_E\sqrt{\alpha_{2_E}}},-\frac{\gamma_D c_{1_E}}{\bar{\gamma}_E},-\frac{\gamma_D c_{2_E}}{\bar{\gamma}_E}\bigg) d\gamma_D.
\end{align}
\par The exact closed-form solution of the integral in (24) is not available in the open literature. Therefore, to solve the above integral, we firstly express $\text{ln}(.)$ in terms of Fox's $H$-function (FHF) by using the property [11, eq. (36)]
\label{eqn_25}
\begin{table*}[h]
\begin{align}
 \text{ln}(1+x)&= H^{1,2}_{2,2} \bigg[ x\bigg\vert
 \begin{matrix}
  (1,1),(1,1)\\
  (1,1),(0,1)
\end{matrix}\bigg] 
\end{align}  
\end{table*}
\par For the confluent Lauricella hypergeometric function $\Phi_2^{(4)}(.)$, the following identity can be utilised [25, 1.ii, pp. 259]   
\label{eqn_26}
\begin{align}
\Phi^{(4)}_2(a_1,a_2,a_3,a_4;b;-x_1 t,-x_2 t,-x_3 t,-x_4 t)= \hspace{6.5 cm} \nonumber\\ \frac{1}{t^{b-1}}\mathcal{L}^{-1} \bigg\{\frac{\Gamma(b)}{s^b} \Big(1+\frac{x_1}{s}\Big)^{-a_1} \Big(1+\frac{x_2}{s}\Big)^{-a_2} \Big(1+\frac{x_3}{s}\Big)^{-a_3} \Big(1+\frac{x_4}{s}\Big)^{-a_4};s, t \bigg \}
\end{align}
\par where $\Re(b)>0$, $\Re(s)>0$, and $\{a_1, a_2, a_3, a_4\} \in \mathbb{R}$ and $\mathcal{L}^{-1}(.)$ is the inverse Laplace transform.
\par To compute inverse Laplace transform, the following identity is recalled [26, eq. (1.43)]
\label{eqn_27}
\begin{table*}[h]
\begin{align}
(1+x)^{-a}= \frac{1}{\Gamma(a)}H^{1,1}_{1,1} \bigg[x\bigg\vert
 \begin{matrix}
  (1-a,1)\\
  (0,1)
\end{matrix}\bigg]
\end{align}
\end{table*}
\par With the aid of (27) and using the definition of FHF that is presented in [26, eq. (1.2) and [26, eq.(1.3)], the inverse Laplace transform of (26) can be rewritten as 
\setcounter{equation}{27}
\label{eqn_28}
\begin{align}
\mathcal{L}^{-1} \bigg\{\frac{\Gamma(b)}{s^b} \Big(1+\frac{x_1}{s}\Big)^{-a_1} \Big(1+\frac{x_2}{s}\Big)^{-a_2} \Big(1+\frac{x_3}{s}\Big)^{-a_3} \Big(1+\frac{x_4}{s}\Big)^{-a_4};s, t \bigg \}=\frac{\Gamma(b)}{t^{b-1} \Gamma(a_1) \Gamma(a_2) \Gamma(a_3) \Gamma(a_4)}   \nonumber\\ 
\frac{1}{(2\pi j)^4} \int_{\mathcal{R}_1} \int_{\mathcal{R}_2}  \int_{\mathcal{R}_3} \int_{\mathcal{R}_4} \Gamma(r_1) \Gamma(a_1-r_1) \Gamma(r_2) \Gamma(a_2-r_2) \Gamma(r_3) \Gamma(a_3-r_3) \Gamma(r_4) \Gamma(a_4-r_4) \nonumber\\ 
x^{-r_1}_1 x^{-r_2}_2 x^{-r_3}_3 x^{-r_4}_4\mathcal{L}^{-1}\Big\{s^{r_1+r_2+r_3+r_4-b} ;s, t \Big \} dr_1 dr_2 dr_3 dr_4
\end{align}
where $j=\sqrt{-1}$ and $\mathcal{R}_i$ for $i \in \{1,2,3,4\}$ represent the suitable closed contours in the complex $r_i$-plane.
\par The inverse Laplace transform in (28) can be calculated as follows 
\label{eqn_29}
\begin{align}
\mathcal{L}^{-1}\Big\{s^{r_1+r_2+r_3+r_4-b} ;s, t \Big \} = \frac{t^{b-r_1-r_2-r_3-r_4-1}}{\Gamma(b-r_1-r_2-r_3-r_4)}
\end{align}
\par Substituting (29) into (28) to yield
\label{eqn_30}
\begin{align}
\mathcal{L}^{-1} \bigg\{\frac{\Gamma(b)}{s^b} \Big(1+\frac{x_1}{s}\Big)^{-a_1} \Big(1+\frac{x_2}{s}\Big)^{-a_2} \Big(1+\frac{x_3}{s}\Big)^{-a_3} \Big(1+\frac{x_4}{s}\Big)^{-a_4};s, t \bigg \}=\frac{\Gamma(b)}{\Gamma(a_1) \Gamma(a_2) \Gamma(a_3) \Gamma(a_4)} t^{b-1} \nonumber\\  
\frac{1}{(2\pi j)^4}  \int_{\mathcal{R}_1} \int_{\mathcal{R}_2}  \int_{\mathcal{R}_3} \int_{\mathcal{R}_4} \frac{\Gamma(r_1) \Gamma(a_1-r_1) \Gamma(r_2) \Gamma(a_2-r_2) \Gamma(r_3) \Gamma(a_3-r_3) \Gamma(r_4) \Gamma(a_4-r_4)}{\Gamma(b-r_1-r_2-r_3-r_4)} \nonumber\\ 
(xt)^{-r_1}_1 (xt)^{-r_2}_2 (xt)^{-r_3}_3 (xt)^{-r_4}_4 dr_1 dr_2 dr_3 dr_4
\end{align}
\par Plugging (25) in (24) and employing (26) and (30) for both confluent Lauricella hypergeometric functions $\Phi_2^{(4)}(.)$, we have 

\setcounter{equation}{30}
\begin{table*}[h]
\label{eqn_31}
\begin{align}
I_1&=\frac{\Omega_D \Omega_E}{[\Gamma(\frac{\mu_D}{2}-m_D)\Gamma(\frac{\mu_E}{2}-m_E)\Gamma(m_D)\Gamma(m_E)]^2}\frac{1}{(2\pi j)^8}  \nonumber\\ 
&\int_{\mathcal{R}_1} \cdots \int_{\mathcal{R}_8} \frac{[\prod_{j=1}^8 \Gamma(r_j)] [\prod_{j=1}^2 \Gamma(\frac{\mu_D}{2}-m_D-r_j) \Gamma(m_D-r_{j+2}) \Gamma(\frac{\mu_E}{2}-m_E-r_{j+4}) \Gamma(m_E-r_{j+6})]}{\Gamma(\mu_D-r_1-\cdots-r_4) \Gamma(1+\mu_E-r_5-\cdots-r_8)} \nonumber\\ 
&\frac{1}{(\bar{\gamma}_D \sqrt{\eta_D \alpha_{2_D}})^{-r_1}} \bigg(\frac{\sqrt{\eta_D}}{\bar{\gamma}_D \sqrt{\alpha_{2_D}}}\bigg)^{-r_2} 
\bigg(\frac{c_{1_D}}{\bar{\gamma}_D}\bigg)^{-r_3} \bigg(\frac{c_{2_D}}{\bar{\gamma}_D} \bigg)^{-r_4} 
\frac{1}{(\bar{\gamma}_E \sqrt{\eta_E \alpha_{2_E}})^{-r_5}} \bigg(\frac{\sqrt{\eta_E}}{\bar{\gamma}_E \sqrt{\alpha_{2_E}}}\bigg)^{-r_6} \nonumber\\ 
&\bigg(\frac{c_{1_E}}{\bar{\gamma}_E}\bigg)^{-r_7} \bigg(\frac{c_{2_E}}{\bar{\gamma}_E} \bigg)^{-r_8}\underbrace{\int_0^\infty \gamma^{\mu_D+\mu_E-\sum_{j=1}^8 r_j-1}_D \text{ln}(1+\gamma_D) d\gamma_D}_{\mathcal{K}_1} dr_1 \cdots dr_8.
\end{align}
\end{table*}
\par With the help of [27, eq. (4.293.10)], the inner integral, $\mathcal{K}_1$, of (31) can be be computed in exact closed-form as   
\label{eqn_32}
\begin{align}
\mathcal{K}_1 &= \int_0^\infty \gamma^{\mu_D+\mu_E-\sum_{j=1}^8 r_j-1}_D \text{ln}(1+\gamma_D) d\gamma_D \nonumber\\
& = \frac{\pi}{(\mu_D+\mu_E-\sum_{j=1}^8 r_j) \text{sin}((\mu_D+\mu_E-\sum_{j=1}^8 r_j) \pi)}
\end{align}
\par Recalling the identities $a=\frac{\Gamma(1+a)}{\Gamma(a)}$ [27, eq. (8.331.1)] and $\text{sin}(\pi a)=\frac{\pi}{a\Gamma(1-a)}$ [27, eq. (8.334.3)], (32) becomes
\label{eqn_33}
\begin{align}
\mathcal{K}_1 = \frac{[\Gamma(\mu_D+\mu_E-\sum_{j=1}^8 r_j)]^2 \Gamma(1-\mu_D-\mu_E+\sum_{j=1}^8 r_j)}{\Gamma(1+\mu_D+\mu_E-\sum_{j=1}^8 r_j)}
\end{align}
\par Inserting (33) in (31) with some mathematical manipulations, (34) is obtained as follows 
\begin{table*}[h]
\label{eqn_34}
\begin{align}
I_1&=\frac{\Omega_D \Omega_E}{[\Gamma(\frac{\mu_D}{2}-m_D)\Gamma(\frac{\mu_E}{2}-m_E)\Gamma(m_D)\Gamma(m_E)]^2} \nonumber\\ 
&\frac{1}{(2\pi j)^8} \int_{\mathcal{R}_1} \cdots \int_{\mathcal{R}_8} \frac{[\Gamma(\mu_D+\mu_E-\sum_{j=1}^8 r_j)]^2 \Gamma(1-\mu_D-\mu_E+\sum_{j=1}^8 r_j) [\prod_{j=1}^8 \Gamma(r_j)] }{\Gamma(1+\mu_D+\mu_E-\sum_{j=1}^8 r_j)} \nonumber\\ 
& \frac{[\prod_{j=1}^2 \Gamma(\frac{\mu_D}{2}-m_D-r_j) \Gamma(m_D-r_{j+2}) \Gamma(\frac{\mu_E}{2}-m_E-r_{j+4}) \Gamma(m_E-r_{j+6})]}{\Gamma(\mu_D-r_1-\cdots-r_4) \Gamma(1+\mu_E-r_5-\cdots-r_8)} \frac{1}{(\bar{\gamma}_D \sqrt{\eta_D \alpha_{2_D}})^{-r_1}}  \nonumber\\ 
&\bigg(\frac{\sqrt{\eta_D}}{\bar{\gamma}_D \sqrt{\alpha_{2_D}}}\bigg)^{-r_2} \bigg(\frac{c_{1_D}}{\bar{\gamma}_D}\bigg)^{-r_3} \bigg(\frac{c_{2_D}}{\bar{\gamma}_D} \bigg)^{-r_4} \frac{1}{(\bar{\gamma}_E \sqrt{\eta_E \alpha_{2_E}})^{-r_5}} \bigg(\frac{\sqrt{\eta_E}}{\bar{\gamma}_E \sqrt{\alpha_{2_E}}}\bigg)^{-r_6} \bigg(\frac{c_{1_E}}{\bar{\gamma}_E}\bigg)^{-r_7} \bigg(\frac{c_{2_E}}{\bar{\gamma}_E} \bigg)^{-r_8} dr_1 \cdots dr_8.
\end{align}
\end{table*}
\par It can be noted that (34) can be written in exact closed-from in terms of the EGBFHF via using [26, eq. (A.1)] and this completes the proof for (10).
\par It can be observed (10) can be utilised to evaluate (11) by replacing each $D$ and $E$ with $E$ and $D$, respectively.
\par For $I_3$, we Substitute (2) in (9) to yield 
\label{eqn_35}
\begin{table*}[h]
\begin{align}
I_3&=\frac{\Omega_E}{[\Gamma(\frac{\mu_E}{2}-m_E)\Gamma(m_E)]^2} \nonumber\\ 
&\frac{1}{(2\pi j)^4} \int_{\mathcal{R}_1} \cdots \int_{\mathcal{R}_4} \frac{[\prod_{j=1}^4 \Gamma(r_j)] [\prod_{j=1}^2 \Gamma(\frac{\mu_E}{2}-m_E-r_{j}) \Gamma(m_E-r_{j+2})]}{\Gamma(\mu_E-r_1-\cdots-r_4)} \nonumber\\ 
&\frac{1}{(\bar{\gamma}_E \sqrt{\eta_E \alpha_{2_E}})^{-r_1}} \bigg(\frac{\sqrt{\eta_E}}{\bar{\gamma}_E \sqrt{\alpha_{2_E}}}\bigg)^{-r_2} \bigg(\frac{c_{1_E}}{\bar{\gamma}_E}\bigg)^{-r_3} \bigg(\frac{c_{2_E}}{\bar{\gamma}_E} \bigg)^{-r_4}
\underbrace{\int_0^\infty \gamma^{\mu_E-\sum_{j=1}^4 r_j-1}_E \text{ln}(1+\gamma_E) d\gamma_E}_{\mathcal{K}_2} dr_1 \cdots dr_4.
\end{align}
\end{table*}
\par Following the same steps of computing $\mathcal{K}_1$ of (31), $\mathcal{K}_2$ is obtained as follows 
\label{eqn_36}
\begin{align}
\mathcal{K}_2=\frac{[\Gamma(\mu_E-\sum_{j=1}^4 r_j)]^2 \Gamma(1-\mu_E+\sum_{j=1}^4 r_j)}{\Gamma(1+\mu_E-\sum_{j=1}^4 r_j)}
\end{align}
\par Substituting (36) in (35), we have 
\begin{table*}[h]
\label{eqn_37}
\begin{align}
I_3&=\frac{\Omega_E}{[\Gamma(\frac{\mu_E}{2}-m_E)\Gamma(m_E)]^2} \nonumber\\ 
&\frac{1}{(2\pi j)^4} \int_{\mathcal{R}_1} \cdots \int_{\mathcal{R}_4} \frac{\Gamma(1-\mu_E+\sum_{j=1}^4 r_j) [\Gamma(\mu_E-\sum_{j=1}^4 r_j)]^2}{\Gamma(1+\mu_E-\sum_{j=1}^4 r_j) [\prod_{j=1}^4 \Gamma(r_j)] } \nonumber\\ 
&\frac{[\prod_{j=1}^2 \Gamma(\frac{\mu_E}{2}-m_E-r_j) \Gamma(m_E-r_{j+2})]}{ \Gamma(\mu_E-\sum_{j=1}^4 r_j)}
\frac{1}{(\bar{\gamma}_E \sqrt{\eta_E \alpha_{2_E}})^{-r_1}} \bigg(\frac{\sqrt{\eta_E}}{\bar{\gamma}_E \sqrt{\alpha_{2_E}}}\bigg)^{-r_2} 
\bigg(\frac{c_{1_E}}{\bar{\gamma}_E}\bigg)^{-r_3} \bigg(\frac{c_{2_E}}{\bar{\gamma}_E} \bigg)^{-r_4} dr_1 \cdots dr_4.
\end{align}
\end{table*}
\par Invoking [26, eq. (A.1)], $I_3$ in (37) is expressed in exact closed-form expression as provided in (12) which completes the proof.

\section*{Appendix B}
\section*{Proof of Corollary 1}
For $\textbf{Case\_2}$, $I_1$ can be calculated via inserting (5) and (6) in (7). Accordingly, this yields 
\label{eqn_38}
\begin{align}
I_1&=\Omega_D \sum_{i_D=1}^{N_D (m_D,\mu_D)} \sum_{j_D=1}^{|\omega_{i_D}|}\frac{A_{{i_D}{j_D}}}{(j_D-1)!} \Bigg[\int_0^\infty \gamma^{j_D-1}_D \text{ln}(1+\gamma_D) e^{-\frac{\vartheta_{i_D}}{\bar{\gamma}_D}\gamma_D} d\gamma_D+ \nonumber \\
&\Omega_E \sum_{i_E=1}^{N_E (m_E,\mu_E)} \sum_{j_E=1}^{|\omega_{i_E}|} 
\frac{B_{{i_E}{j_E}}}{(j_E-1)!} \int_0^\infty \gamma^{j_D+j_E-2}_D \text{ln}(1+\gamma_D) e^{-\big(\frac{\vartheta_{i_D}}{\bar{\gamma}_D}+\frac{\vartheta_{i_E}}{\bar{\gamma}_E}\big)\gamma_D} d\gamma_D \Bigg].
\end{align}
\par Both integrals of (38) can be computed in simple exact closed-form expressions with the aid of [28, eq. (47)] 
\label{eqn_39}
\begin{align}
\int_0^\infty x^{a-1} \text{ln}(1+x) e^{-bx} dx =\Gamma(a) e^b \sum_{k=1}^a \frac{\Gamma(k-a, b)}{b^k}
\end{align}
\par Using (39) and doing some mathematical simplifications, $I_1$ that is given in (13) is deduced and this completes the proof.  
\par Using $D$ and $E$ instead of $E$ and $D$, respectively, in (13), the result is $I_2$ for integer $\mu_l$ and $m_l$ as shown in (14).
\par Similarly, after plugging (5) in (9) and using [28, eq. (47)], $I_3$ is expressed in exact closed-form as given in (15) which completes the proof.
\section*{Appendix C}
\section*{Proof of Theorem 2}
\par Inserting (2) and (4) in (17), the result is
\label{eqn_40}
\begin{align}
\text{SOP}&=\frac{\Omega_D \Omega_E}{\Gamma(\mu_E) \Gamma(\mu_D+1)}\int_0^\infty (\theta \gamma_E+\theta-1)^{\mu_D} \gamma^{\mu_E-1}_E 
\nonumber\\
&\times \Phi^{(4)}_2 \bigg(\frac{\mu_E}{2}-m_E,\frac{\mu_E}{2}-m_E,m_E,m_E;\mu_E;-\frac{\gamma_E}{\bar{\gamma}_E\sqrt{\eta_E \alpha_{2_E}}},-\frac{\gamma_E \sqrt{\eta_E}}{\bar{\gamma}_E\sqrt{\alpha_{2_E}}},-\frac{\gamma_E c_{1_E}}{\bar{\gamma}_E},-\frac{\gamma_E c_{2_E}}{\bar{\gamma}_E}\bigg)
\nonumber\\
&\times \Phi^{(4)}_2 \bigg(\frac{\mu_D}{2}-m_D,\frac{\mu_D}{2}-m_D,m_D,m_D;\mu_D+1;-\frac{(\theta \gamma_E+\theta-1)}{\bar{\gamma}_D\sqrt{\eta_D \alpha_{2_D}}},-\frac{(\theta \gamma_E+\theta-1) \sqrt{\eta_D}}{\bar{\gamma}_D\sqrt{\alpha_{2_D}}},\nonumber\\
&-\frac{(\theta \gamma_E+\theta-1) c_{1_D}}{\bar{\gamma}_D},-\frac{(\theta \gamma_E+\theta-1) c_{2_D}}{\bar{\gamma}_D}\bigg) d\gamma_E.
\end{align}
\par To compute the integral in (40), we recall (26) and (30) for both confluent Lauricella hypergeometric functions $\Phi_2^{(4)}(.)$. Hence, we have
\label{eqn_41}
\begin{table*}[h]
\begin{align}
\text{SOP}&=\frac{\Omega_D \Omega_E}{[\Gamma(\frac{\mu_D}{2}-m_D)\Gamma(\frac{\mu_E}{2}-m_E)\Gamma(m_D)\Gamma(m_E)]^2}\frac{1}{(2\pi j)^8}  \nonumber\\ 
&\int_{\mathcal{R}_1} \cdots \int_{\mathcal{R}_8} \frac{[\prod_{j=1}^8 \Gamma(r_j)] [\prod_{j=1}^2 \Gamma(\frac{\mu_E}{2}-m_E-r_j) \Gamma(m_E-r_{j+2}) \Gamma(\frac{\mu_D}{2}-m_D-r_{j+4}) \Gamma(m_D-r_{j+6})]}{\Gamma(1+\mu_E-r_1-\cdots-r_4) \Gamma(\mu_D-r_5-\cdots-r_8)} \nonumber\\ 
&\frac{1}{(\bar{\gamma}_E \sqrt{\eta_E \alpha_{2_E}})^{-r_1}} \bigg(\frac{\sqrt{\eta_E}}{\bar{\gamma}_E \sqrt{\alpha_{2_E}}}\bigg)^{-r_2} \bigg(\frac{c_{1_E}}{\bar{\gamma}_E}\bigg)^{-r_3} \bigg(\frac{c_{2_E}}{\bar{\gamma}_E} \bigg)^{-r_4}\frac{1}{(\bar{\gamma}_D \sqrt{\eta_D \alpha_{2_D}})^{-r_5}} \bigg(\frac{\sqrt{\eta_D}}{\bar{\gamma}_D \sqrt{\alpha_{2_D}}}\bigg)^{-r_6} 
\nonumber\\ 
&\bigg(\frac{c_{1_D}}{\bar{\gamma}_D}\bigg)^{-r_7} \bigg(\frac{c_{2_D}}{\bar{\gamma}_D} \bigg)^{-r_8}
\underbrace{\int_0^\infty   \gamma^{\mu_E-\sum_{j=1}^4 r_j -1}_E (\theta \gamma_E +\theta-1)^{\mu_D-\sum_{j=5}^8 r_j} d\gamma_E}_{\mathcal{K}_3} dr_1 \cdots dr_8.
\end{align}
\end{table*}   
\par Performing some mathematical manipulations and using [27, eq. (3.194.3)], $\mathcal{K}_3$ of (41) can be evaluated in exact closed-form as follows  
\label{eqn_42}
\begin{align}
\mathcal{K}_3& = \theta^{-\mu_E+\sum_{j=1}^4 r_j}  (\theta-1)^{\mu_D+\mu_E-\sum_{j=1}^8 r_j} B(\mu_E-\textstyle{\sum\nolimits_{j=1}^4} r_j, -\mu_D -\mu_E+\textstyle{\sum\nolimits_{j=1}^8} r_j ) 
\end{align}
\par where $B(.,.)$ is the Beta function defined in [25, eq. (1.1.43)]. 
\par Invoking the identity $B(x,y)=\frac{\Gamma(x) \Gamma(y)}{\Gamma(x+y)}$ [25, eq. (1.1.47)], (42) can be rewritten as 
\label{eqn_43}
\begin{align}
\mathcal{K}_3& = \frac{\Gamma(\mu_E -\textstyle{\sum\nolimits_{j=1}^4} r_j) \Gamma(-\mu_D-\mu_E+ \textstyle{\sum\nolimits_{j=1}^8} r_j)}{\Gamma(-\mu_D+\textstyle{\sum\nolimits_{j=5}^8} r_j)}  \theta^{-\mu_E+\sum_{j=1}^4 r_j} (\theta-1)^{\mu_D+\mu_E-\sum_{j=1}^8 r_j} 
\end{align}
\par Plugging (43) in (41) to yield (44) that is given at the top of the next page 
 \label{eqn_44}
\begin{table*}[h]
\begin{align}
\text{SOP}&=\frac{\Omega_D \Omega_E}{[\Gamma(\frac{\mu_D}{2}-m_D)\Gamma(\frac{\mu_E}{2}-m_E)\Gamma(m_D)\Gamma(m_E)]^2} \frac{(\theta-1)^{\mu_D+\mu_E}}{\theta^{\mu_E}}\frac{1}{(2\pi j)^8} \nonumber\\ 
&\int_{\mathcal{R}_1} \cdots \int_{\mathcal{R}_8} \frac{[\prod_{j=1}^8 \Gamma(r_j)] [\prod_{j=1}^2 \Gamma(\frac{\mu_E}{2}-m_E-r_j) \Gamma(m_E-r_{j+2}) \Gamma(\frac{\mu_D}{2}-m_D-r_{j+4}) \Gamma(m_D-r_{j+6})]}{\Gamma(1+\mu_E-r_1-\cdots-r_4) \Gamma(\mu_D-r_5-\cdots-r_8)} \nonumber\\ 
&\frac{\Gamma(\mu_E -\textstyle{\sum\nolimits_{j=1}^4} r_j) \Gamma(-\mu_D-\mu_E+ \textstyle{\sum\nolimits_{j=1}^8} r_j)}{\Gamma(-\mu_D+\textstyle{\sum\nolimits_{j=5}^8} r_j)} \bigg(\frac{(\theta-1)}{\theta \bar{\gamma}_E \sqrt{\eta_E \alpha_{2_E}}}\bigg)^{-r_1} \bigg(\frac{(\theta-1)\sqrt{\eta_E}}{\theta \bar{\gamma}_E \sqrt{\alpha_{2_E}}}\bigg)^{-r_2} \nonumber\\ 
&\bigg(\frac{(\theta-1) c_{1_E}}{\theta \bar{\gamma}_E}\bigg)^{-r_3} \bigg(\frac{(\theta-1) c_{2_E}}{\theta \bar{\gamma}_E} \bigg)^{-r_4}\bigg(\frac{(\theta-1)}{\bar{\gamma}_D \sqrt{\eta_D \alpha_{2_D}}}\bigg)^{-r_5} \bigg(\frac{(\theta-1)\sqrt{\eta_D}}{\bar{\gamma}_D \sqrt{\alpha_{2_D}}}\bigg)^{-r_6} \bigg(\frac{(\theta-1) c_{1_D}}{\bar{\gamma}_D}\bigg)^{-r_7}
\nonumber\\ 
&  \bigg(\frac{(\theta-1) c_{2_D}}{\bar{\gamma}_D} \bigg)^{-r_8} dr_1 \cdots dr_8.
\end{align}
\hrulefill
\vspace*{1pt} 
\end{table*}
\par Again, [26, eq. (A.1)] is utilised for (44) to complete the proof of (18). 
\par The SOP for $\textbf{Case\_2}$ can calculated after substituting (5) and (6) in (17) and using the fact $\int_0^\infty f_\gamma(\gamma) d\gamma \triangleq 1$. Consequently, we have
\label{eqn_45}
\begin{align}
\text{SOP}&=1+\Omega_D \Omega_E  \sum_{i_E=1}^{N_E (m_E,\mu_E)} \sum_{j_E=1}^{|\omega_{i_E}|} \frac{A_{{i_E}{j_E}}}{(j_E-1)!} \sum_{i_D=1}^{N_D (m_D,\mu_D)} \sum_{j_D=1}^{|\omega_{i_D}|} \frac{B_{{i_D}{j_D}}}{(j_D-1)!} \nonumber\\
&\underbrace{\int_0^\infty \gamma^{j_E-1}_E (\theta \gamma_E+\theta-1)^{j_D-1} e^{-\frac{\vartheta_{i_E}}{\bar{\gamma}_E} \gamma_E-(\theta \gamma_E+\theta-1) \frac{\vartheta_{i_D}}{\bar{\gamma}_D}} d\gamma_E}_{\mathcal{K}_4}
\end{align}
\par Doing some mathematical manipulations on $\mathcal{K}_4$ of (45) to yield 
\label{eqn_46}
\begin{align}
\mathcal{K}_4= (\theta-1)^{j_D-1} e^{-(\theta-1)\frac{\vartheta_{i_D}}{\bar{\gamma}_D}} \int_0^\infty \gamma^{j_E-1}_E \bigg(1+\frac{\theta}{\theta-1} \gamma_E \bigg)^{j_D-1} e^{-\big(\frac{\vartheta_{i_E}}{\bar{\gamma}_E}+\frac{\theta \vartheta_{i_D}}{\bar{\gamma}_D}\big)\gamma_E} d\gamma_E
\end{align}
\par Employing the identity $(1+a)^b=\sum_{k=0}^{b}{{b}\choose{k}}a^k$ [27, eq. (1.111)] in (46), we obtain
\label{eqn_47}
\begin{align}
\mathcal{K}_4= (\theta-1)^{j_D-1} e^{-(\theta-1)\frac{\vartheta_{i_D}}{\bar{\gamma}_D}} \sum_{r=0}^{j_D-1} {{j_D-1}\choose{r}} \bigg(\frac{\theta}{\theta-1}\bigg)^r\int_0^\infty \gamma^{j_E+r-1}_E  e^{-\big(\frac{\vartheta_{i_E}}{\bar{\gamma}_E}+\frac{\theta \vartheta_{i_D}}{\bar{\gamma}_D}\big)\gamma_E} d\gamma_E
\end{align}
\par With the aid of [27, eq. (3.381.4)], the integral in (47) can be evaluated as follows 
\label{eqn_48}
\begin{align}
\mathcal{K}_4= (\theta-1)^{j_D-1} e^{-(\theta-1)\frac{\vartheta_{i_D}}{\bar{\gamma}_D}} \sum_{r=0}^{j_D-1} {{j_D-1}\choose{r}} \bigg(\frac{\theta}{\theta-1}\bigg)^r \frac{\Gamma(j_E+r)}{\vast(\frac{\vartheta_{i_E}}{\bar{\gamma}_E}+\frac{\theta \vartheta_{i_D}}{\bar{\gamma}_D}\vast)^{j_E+r}}
\end{align}
\par Plugging (48) in (45) with some mathematical simplifications, the result is (19) which completes the proof. 
\section*{Appendix D}
\section*{Proof of Theorem 3}
\par Substituting (2) and (4) in (20), this yields
\label{eqn_49}
\begin{align}
\text{SOP}^L&=\frac{\Omega_D \Omega_E}{\Gamma(\mu_E) \Gamma(\mu_D+1)} \theta^{\mu_D} \int_0^\infty  \gamma^{\mu_E+\mu_D-1}_E 
\nonumber\\
&\times \Phi^{(4)}_2 \bigg(\frac{\mu_E}{2}-m_E,\frac{\mu_E}{2}-m_E,m_E,m_E;\mu_E;-\frac{\gamma_E}{\bar{\gamma}_E\sqrt{\eta_E \alpha_{2_E}}},-\frac{\sqrt{\eta_E}}{\bar{\gamma}_E\sqrt{\alpha_{2_E}}} \gamma_E ,-\frac{ c_{1_E}}{\bar{\gamma}_E} \gamma_E,-\frac{ c_{2_E}}{\bar{\gamma}_E}\gamma_E\bigg)
\nonumber\\
&\times \Phi^{(4)}_2 \bigg(\frac{\mu_D}{2}-m_D,\frac{\mu_D}{2}-m_D,m_D,m_D;\mu_D+1;-\frac{\theta }{\bar{\gamma}_D\sqrt{\eta_D \alpha_{2_D}}} \gamma_E,-\frac{\theta \sqrt{\eta_D}}{\bar{\gamma}_D\sqrt{\alpha_{2_D}}} \gamma_E,\nonumber\\
&-\frac{\theta c_{1_D}}{\bar{\gamma}_D} \gamma_E ,-\frac{ \theta c_{2_D}}{\bar{\gamma}_D}  \gamma_E\bigg) d\gamma_E.
\end{align}
\par To calculate the integral in (49), the following property can be used [29, p. 177]
\label{eqn_50}
\begin{align}
e^{-c_i}\Phi^{(n)}_2\big(a_1,\cdots,a_n;b;c_1,\cdots,c_n\big)= 
\Phi^{(n)}_2\big(a_1,.,a_{i-1},b-a_1-.-a_n,a_{i+1},.,a_n;b;c_1-c_i,\cdots, \nonumber\\
c_{i-1}-c_i,-c_i,c_{i+1}-c_i,\cdots,c_n-c_i\big)
\end{align}
Accordingly, both confluent Lauricella hypergeometric functions $\Phi_2^{(4)}(.)$ become 
\label{eqn_51}
\begin{align}
\Phi^{(4)}_2 \bigg(\frac{\mu_E}{2}-m_E,\frac{\mu_E}{2}-m_E,m_E,m_E;\mu_E;-\frac{\gamma_E}{\bar{\gamma}_E\sqrt{\eta_E \alpha_{2_E}}},-\frac{\sqrt{\eta_E}}{\bar{\gamma}_E\sqrt{\alpha_{2_E}}} \gamma_E ,-\frac{ c_{1_E}}{\bar{\gamma}_E} \gamma_E,-\frac{ c_{2_E}}{\bar{\gamma}_E}\gamma_E\bigg)\nonumber\\ 
= e^{-\frac{\gamma_E}{\bar{\gamma}_E \sqrt{\eta_E \alpha_{2_E}}}} \Phi^{(4)}_2 \bigg(0,\frac{\mu_E}{2}-m_E,m_E,m_E;\mu_E;\frac{\gamma_E}{\bar{\gamma}_E\sqrt{\eta_E \alpha_{2_E}}},\frac{1-\eta_E }{\bar{\gamma}_E \sqrt{\eta_E \alpha_{2_E}}} \gamma_E,\nonumber\\
\frac{1- \sqrt{\eta_E \alpha_{2_E}} c_{1_E}}{\bar{\gamma}_E \sqrt{\eta_E \alpha_{2_E}}} \gamma_E ,\frac{1- \sqrt{\eta_E \alpha_{2_E}} c_{2_E}}{\bar{\gamma}_E \sqrt{\eta_E \alpha_{2_E}}} \gamma_E \bigg)
\end{align}
and
\label{eqn_52}
\begin{align}
\Phi^{(4)}_2 \bigg(\frac{\mu_D}{2}-m_D,\frac{\mu_D}{2}-m_D,m_D,m_D;\mu_D+1;-\frac{\theta }{\bar{\gamma}_D \sqrt{\eta_D \alpha_{2_D}}} \gamma_E,-\frac{\theta \sqrt{\eta_D}}{\bar{\gamma}_D\sqrt{\alpha_{2_D}}} \gamma_E ,-\frac{ \theta c_{1_D}}{\bar{\gamma}_D} \gamma_E,-\frac{ \theta c_{2_D}}{\bar{\gamma}_D}\gamma_E\bigg)\nonumber\\ 
= e^{-\frac{\theta}{\bar{\gamma}_D \sqrt{\eta_D \alpha_{2_D}}}  \gamma_E} \Phi^{(4)}_2 \bigg(0,\frac{\mu_D}{2}-m_D,m_D,m_D;\mu_D+1;\frac{\theta }{\bar{\gamma}_D \sqrt{\eta_D \alpha_{2_D}}} \gamma_E,\frac{(1-\eta_D) \theta}{\bar{\gamma}_D \sqrt{\eta_D \alpha_{2_D}}} \gamma_E,\nonumber\\
\frac{(1- \sqrt{\eta_D \alpha_{2_D}} c_{1_D}) \theta}{\bar{\gamma}_D \sqrt{\eta_D \alpha_{2_D}}} \gamma_E ,\frac{(1- \sqrt{\eta_D \alpha_{2_D}} c_{2_D}) \theta}{\bar{\gamma}_D \sqrt{\eta_D \alpha_{2_D}}} \gamma_E \bigg)
\end{align}
\par Inserting (51) and (52) in (49), we have (53) at the top of the next page.
\begin{table*}[h]
\label{eqn_53}
\begin{align}
\text{SOP}^L&=\frac{\Omega_D \Omega_E}{\Gamma(\mu_E) \Gamma(\mu_D+1)} \theta^{\mu_D} \int_0^\infty  \gamma^{\mu_E+\mu_D-1}_E e^{-\big(\frac{1}{\bar{\gamma}_E \sqrt{\eta_E \alpha_{2_E}}}+\frac{\theta}{\bar{\gamma}_D \sqrt{\eta_D \alpha_{2_D}}}\big)\gamma_E }
\nonumber\\
&\times \Phi^{(4)}_2 \bigg(0,\frac{\mu_E}{2}-m_E,m_E,m_E;\mu_E;\frac{\gamma_E}{\bar{\gamma}_E\sqrt{\eta_E \alpha_{2_E}}},
\frac{1-\eta_E }{\bar{\gamma}_E \sqrt{\eta_E \alpha_{2_E}}} \gamma_E,
\nonumber\\
&\frac{1-\sqrt{\eta_E \alpha_{2_E}} c_{1_E}}{\bar{\gamma}_E \sqrt{\eta_E \alpha_{2_E}}} \gamma_E ,
\frac{1- \sqrt{\eta_E \alpha_{2_E}} c_{2_E}}{\bar{\gamma}_E \sqrt{\eta_E \alpha_{2_E}}} \gamma_E \bigg)
\nonumber\\
&\times  \Phi^{(4)}_2 \bigg(0,\frac{\mu_D}{2}-m_D,m_D,m_D;\mu_D+1;
\frac{\theta }{\bar{\gamma}_D \sqrt{\eta_D \alpha_{2_D}}} \gamma_E,
\frac{(1-\eta_D) \theta}{\bar{\gamma}_D \sqrt{\eta_D \alpha_{2_D}}} \gamma_E,\nonumber\\
&\frac{(1- \sqrt{\eta_D \alpha_{2_D}} c_{1_D}) \theta}{\bar{\gamma}_D \sqrt{\eta_D \alpha_{2_D}}} \gamma_E ,\frac{(1- \sqrt{\eta_D \alpha_{2_D}} c_{2_D}) \theta}{\bar{\gamma}_D \sqrt{\eta_D \alpha_{2_D}}} \gamma_E \bigg) d\gamma_E.
\end{align}
\hrulefill
\vspace*{1pt} 
\end{table*}
\par Recalling (26) and (30) for both $\Phi^{(4)}_2(.)$ of (53) to yield (54) at the top of the next page
\begin{table*}[h]
\label{eqn_54}
\begin{align}
\text{SOP}^L&=\frac{\Omega_D \Omega_E}{\Gamma(\frac{\mu_D}{2}-m_D)\Gamma(\frac{\mu_E}{2}-m_E)[\Gamma(m_D)\Gamma(m_E)]^2} \theta^\mu \nonumber\\ 
&\frac{1}{(2\pi j)^6} \int_{\mathcal{R}_1} \cdots \int_{\mathcal{R}_6} \frac{[\prod_{j=1}^6 \Gamma(r_j)] \Gamma(\frac{\mu_E}{2}-m_E-r_1)   \Gamma(\frac{\mu_D}{2}-m_D-r_4) [\prod_{j=2}^3 \Gamma(m_E-r_j)\Gamma(m_D-r_{j+3})]}{\Gamma(\mu_E-r_1-\cdots-r_3) \Gamma(1+\mu_D-r_4-\cdots-r_6)} \nonumber\\ 
& \bigg(\frac{\sqrt{\eta_E \alpha_{2_E}} c_{1_E}-1}{\bar{\gamma}_E \sqrt{\eta_E \alpha_{2_E}}}\bigg)^{-r_1} 
\bigg(\frac{\sqrt{\eta_E \alpha_{2_E}} c_{1_E}-1}{\bar{\gamma}_E \sqrt{\eta_E \alpha_{2_E}}} \bigg)^{-r_2} 
\bigg(\frac{\sqrt{\eta_E \alpha_{2_E}} c_{2_E}-1}{\bar{\gamma}_E \sqrt{\eta_E \alpha_{2_E}}} \bigg)^{-r_3}  \nonumber\\
& \bigg(\frac{(\eta_D-1) \theta}{\bar{\gamma}_D \sqrt{\eta_D \alpha_{2_D}}}\bigg)^{-r_4} 
\bigg(\frac{(\sqrt{\eta_D \alpha_{2_D}} c_{1_D}-1) \theta}{\bar{\gamma}_D \sqrt{\eta_D \alpha_{2_D}}}\bigg)^{-r_5} 
\bigg(\frac{(\sqrt{\eta_D \alpha_{2_D}} c_{2_D}-1) \theta}{\bar{\gamma}_D \sqrt{\eta_D \alpha_{2_D}}}\bigg)^{-r_6}
\nonumber\\ 
&\underbrace{\int_0^\infty \gamma^{\mu_D+\mu_E-\sum_{j=1}^6 r_j-1}_E e^{-\big(\frac{1}{\bar{\gamma}_E \sqrt{\eta_E \alpha_{2_E}}}+\frac{\theta}{\bar{\gamma}_D \sqrt{\eta_D \alpha_{2_D}}}\big)\gamma_E } d\gamma_E}_{\mathcal{K}_5} dr_1 \cdots dr_6.
\end{align}
\end{table*}
\par With the aid of [27, eq. (3.381.4)], $\mathcal{K}_5$ of (54) can be computed as follows
\label{eqn_55}
\begin{align}
\mathcal{K}_5= \frac{\Gamma\big(\mu_D+\mu_E-\sum_{j=1}^6 r_j\big)}{\phi^{\mu_D+\mu_E-\sum_{j=1}^6 r_j}}
\end{align}
\par where $\phi=\frac{1}{\bar{\gamma}_E \sqrt{\eta_E \alpha_{2_E}}}+\frac{\theta}{\bar{\gamma}_D \sqrt{\eta_D \alpha_{2_D}}}$.
\par Plugging (55) in (56) and doing some mathematical operations, (56) is yielded as shown at the top of the next page. 

\begin{table*}[h]
\label{eqn_56}
\begin{align}
\text{SOP}^L&=\frac{\Omega_D \Omega_E \theta^\mu}{\phi^{\mu_D+\mu_E} \Gamma(\frac{\mu_D}{2}-m_D)\Gamma(\frac{\mu_E}{2}-m_E)[\Gamma(m_D)\Gamma(m_E)]^2}  \nonumber\\ 
&\frac{1}{(2\pi j)^6} \int_{\mathcal{R}_1} \cdots \int_{\mathcal{R}_6} \frac{\Gamma\big(\mu_D+\mu_E-\sum_{j=1}^6 r_j\big) [\prod_{j=1}^6 \Gamma(r_j)] \Gamma(\frac{\mu_E}{2}-m_E-r_1)\Gamma(\frac{\mu_D}{2}-m_D-r_4)}{\Gamma(\mu_E-r_1-\cdots-r_3) }\nonumber\\ 
& \frac{[\prod_{j=2}^3 \Gamma(m_E-r_j)\Gamma(m_D-r_{j+3})]}{\Gamma(1+\mu_D-r_4-\cdots-r_6)} \bigg(\frac{\sqrt{\eta_E \alpha_{2_E}} c_{1_E}-1}{\phi \bar{\gamma}_E \sqrt{\eta_E \alpha_{2_E}}}\bigg)^{-r_1} 
\bigg(\frac{\sqrt{\eta_E \alpha_{2_E}} c_{1_E}-1}{\phi \bar{\gamma}_E \sqrt{\eta_E \alpha_{2_E}}} \bigg)^{-r_2} 
  \nonumber\\
& \bigg(\frac{\sqrt{\eta_E \alpha_{2_E}} c_{2_E}-1}{\phi \bar{\gamma}_E \sqrt{\eta_E \alpha_{2_E}}} \bigg)^{-r_3} 
\bigg(\frac{(\eta_D-1) \theta}{\phi \bar{\gamma}_D \sqrt{\eta_D \alpha_{2_D}}}\bigg)^{-r_4} 
\bigg(\frac{(\sqrt{\eta_D \alpha_{2_D}} c_{1_D}-1) \theta}{\phi \bar{\gamma}_D \sqrt{\eta_D \alpha_{2_D}}}\bigg)^{-r_5} 
\bigg(\frac{(\sqrt{\eta_D \alpha_{2_D}} c_{2_D}-1) \theta}{\phi\bar{\gamma}_D \sqrt{\eta_D \alpha_{2_D}}}\bigg)^{-r_6}
\nonumber\\ 
& dr_1 \cdots dr_6.
\end{align}
\hrulefill
\vspace*{1pt} 
\end{table*} 
\par With the help of [26, eq. (A.1)], the SOP$^L$ of (56) is expressed in exact closed-form as given in (21) and this completes the proof.
\par For $\textbf{Case\_2}$, SOP$^L$ can be computed by plugging (5) and (6) in (20) and utilising $\int_0^\infty f_\gamma(\gamma) d\gamma \triangleq 1$. Accordingly, this yields
\label{eqn_57}
\begin{align}
\text{SOP}^L&=1+\Omega_D \Omega_E  \sum_{i_E=1}^{N_E (m_E,\mu_E)} \sum_{j_E=1}^{|\omega_{i_E}|} \frac{A_{{i_E}{j_E}}}{(j_E-1)!} \sum_{i_D=1}^{N_D (m_D,\mu_D)} \sum_{j_D=1}^{|\omega_{i_D}|} \frac{B_{{i_D}{j_D}}}{(j_D-1)!} \nonumber\\
&\theta^{j_D-1} \underbrace{\int_0^\infty \gamma^{j_D+j_E-2}_E e^{-\big(\frac{\vartheta_{i_E}}{\bar{\gamma}_E}+\frac{\theta \vartheta_{i_D}}{\bar{\gamma}_D}\big) \gamma_E} d\gamma_E}_{\mathcal{K}_6}
\end{align}
\par The above integral, $\mathcal{K}_6$, can be computed by using [27, eq. (3.381.4)] as follows
\label{eqn_42}
\begin{align}
\mathcal{K}_6&= \frac{\Gamma(j_D+j_E-1)}{\vast(\frac{\vartheta_{i_E}}{\bar{\gamma}_E}+\frac{\theta \vartheta_{i_D}}{\bar{\gamma}_D}\vast)^{j_D+j_E-1}}
\end{align}

\par Substituting (58) in (57), the result is (22) and the proof is completed

\end{document}